\documentclass[twocolumn]{aastex63}
\usepackage{graphicx}
\usepackage{amsmath}
\usepackage{natbib}
\usepackage{multirow}
\usepackage{tabularx}
\usepackage{graphicx}
\usepackage{lineno}

\received{Feb 5, 2021}
\revised{March 25, 2021}
\accepted{April 1, 2021}
\submitjournal{ApJ}
\shorttitle{The Circumnuclear Disk Revealed by ALMA}
\shortauthors{Hsieh et al.}
\graphicspath{{./}{figures/}}

\begin{document}
\title{The Circumnuclear Disk Revealed by ALMA. I. Dense Clouds and Tides in the Galactic Center}

\author[0000-0001-9155-3978]{Pei-Ying Hsieh}
\affiliation{Joint ALMA Observatory, Alonso de C\'{o}rdova, 3107, Vitacura, Santiago 763-0355, Chile}
\affiliation{European Southern Observatory, Alonso de C\'{o}rdova, 3107, Vitacura, Santiago 763-0355, Chile}
\affiliation{Academia Sinica Institute of Astronomy and Astrophysics, P.O. Box 23-141, Taipei 10617, Taiwan, R.O.C.}
\email{pei-ying.hsieh@alma.cl}

\author[0000-0003-2777-5861]{Patrick M. Koch}
\affiliation{Academia Sinica Institute of Astronomy and Astrophysics, P.O. Box 23-141, Taipei 10617, Taiwan, R.O.C.}

\author[0000-0003-4625-229X]{Woong-Tae Kim}
\affiliation{Department of Physics \& Astronomy, Seoul National University, 1 Gwanak-ro, Gwanak-gu, Seoul 08826, Republic of Korea}

\author[0000-0001-9281-2919]{Sergio Mart\'{i}n}
\affiliation{Joint ALMA Observatory, Alonso de C\'{o}rdova, 3107, Vitacura, Santiago 763-0355, Chile}
\affiliation{European Southern Observatory, Alonso de C\'{o}rdova, 3107, Vitacura, Santiago 763-0355, Chile}

\author[0000-0003-1412-893X]{Hsi-Wei Yen}
\affiliation{Academia Sinica Institute of Astronomy and Astrophysics, P.O. Box 23-141, Taipei 10617, Taiwan, R.O.C.}

\author[0000-0003-2251-0602]{John Carpenter}
\affiliation{Joint ALMA Observatory, Alonso de C\'{o}rdova, 3107, Vitacura, Santiago 763-0355, Chile}

\author[0000-0002-6824-6627]{Nanase Harada}
\affiliation{National Astronomical Observatory of Japan, 2-21-1 Osawa, Mitaka, Tokyo 181-8588, Japan}

\author[0000-0003-2261-5746]{Jean L. Turner}
\affiliation{Department of Physics and Astronomy, UCLA, Los Angeles, CA 90095-1547}

\author[0000-0002-3412-4306]{Paul T. P. Ho}
\affiliation{Academia Sinica Institute of Astronomy and Astrophysics, P.O. Box 23-141, Taipei 10617, Taiwan, R.O.C.}
\affiliation{East Asian Observatory, 660 N. Aohoku Place, University Park, Hilo, Hawaii 96720, U.S.A.}

\author[0000-0002-0675-276X]{Ya-Wen Tang}
\affiliation{Academia Sinica Institute of Astronomy and Astrophysics, P.O. Box 23-141, Taipei 10617, Taiwan, R.O.C.}

\author[0000-0002-5770-8494]{Sara C. Beck}
\affiliation{School of Physics and Astronomy, Tel Aviv University, Ramat Aviv, Israel}

\begin{abstract}
Utilizing the Atacama Large Millimeter/submillimeter Array (ALMA), we present  CS line maps in five rotational lines ($J_{\rm u}=7,5,4,3,2$) toward the circumnuclear disk (CND) and streamers of the Galactic Center.
Our primary goal is to resolve the compact structures within the CND and the streamers, in order to understand the stability conditions of molecular cores in the vicinity of the supermassive black hole (SMBH) Sgr A*. Our data provide the first homogeneous high-resolution (1.3$\arcsec$ = 0.05 pc) observations aiming at resolving  density and temperature structures. The  CS  clouds have sizes of 0.05--0.2 pc  with a broad range of velocity dispersion ($\sigma_{\rm FWHM}=5-40$ km s$^{-1}$). The CS clouds are a mixture of warm ($T_{\rm k}\ge 50-500$ K, n$_{\rm H_2}$=$10^{3-5}$ cm$^{-3}$) and cold gas ($T_{\rm k}\le 50$ K, n$_{\rm H_2}$=$10^{6-8}$ cm$^{-3}$). A stability analysis based on the unmagnetized virial theorem including tidal force shows that $84${\raisebox{0.5ex}{\tiny$\substack{+16 \\ -37}$}} \% of the total gas mass is tidally stable, which accounts for the majority of gas mass. Turbulence dominates the internal energy and thereby sets the threshold densities 10--100 times higher than the tidal limit at distance $\ge$  1.5 pc to Sgr A*, and therefore, inhibits the clouds from collapsing to form stars near the SMBH. However, within the central 1.5 pc, the tidal force overrides turbulence and the threshold densities for a gravitational collapse quickly grow to $\ge 10^{8}$ cm$^{-3}$.
\end{abstract}


\section{Introduction} \label{sec:intro}
\subsection{Molecular Inflows in the Galactic Center}
\begin{figure*}[ht!]
\epsscale{1.15}
\plotone{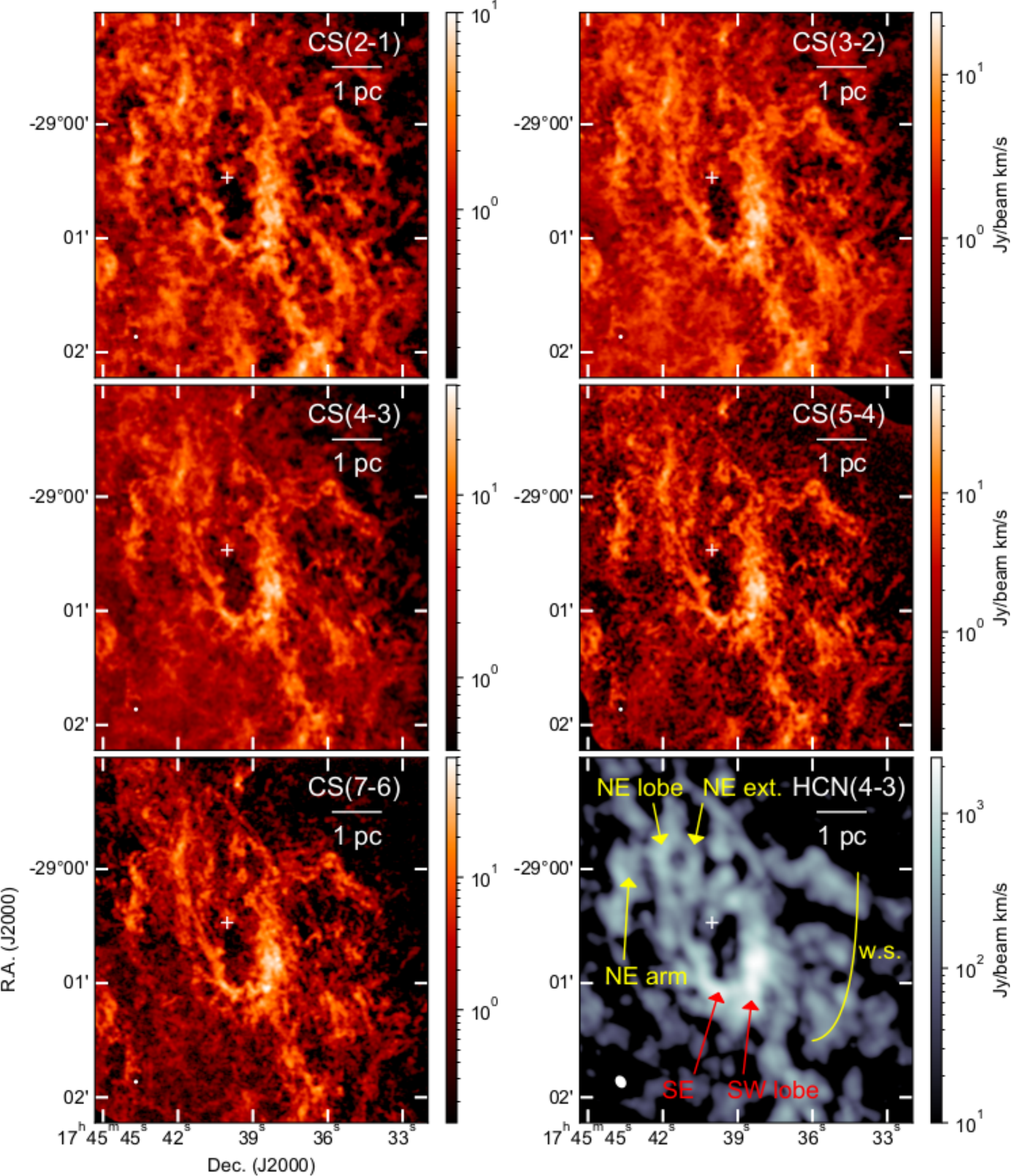}
\caption{Upper: ALMA integrated intensity maps of CS(2-1), CS(3-2), CS(4-3), CS(5-4), and CS(7-6) tracing the CND and the streamers.
Nomenclature of previously identified features in \citet{chris,maria,martin12} are labeled on the SMA HCN(4-3) map \citep{liu12}. The location of Sgr A* is labeled with the cross.  The beam sizes are shown with the white ellipse (lower left corner), which are 1.3$\arcsec$ and 5.8$\arcsec\times4.4\arcsec$ for the ALMA and SMA maps, respectively. Scale bar of 1 pc is labeled. \label{fig-m0}}
\end{figure*}

The mechanisms that transport molecular inflows from kilo-pc down to a few pc are a key problem in galaxy evolution, with an ultimate goal of understanding  the co-evolution of inflows, star formation, and the active cycle of a supermassive black hole (SMBH) in a galaxy  \citep[e.g.][]{collin99, hopkins10,kim12}. The gravitational torque exerted on the gas by a stellar bar has often been invoked to drive the inflows onto a kilo-pc and hundred-pc disk or ring \citep[e.g.,][]{bar92,sakamoto99,hsieh11,lin13,sormani19,tress20}. Within the central few pc, previous observations show that the concentration of molecular gas is influenced by the gravitational potential of a SMBH and follows a quasi-Keplerian rotation, which is often called the circumnuclear disk/ring \citep[CND/CNR, e.g.][]{martin12,liu12, mills13b,hsieh17,imanishi18,tsuboi18,javier18,izumi18,combes19}.
The CND/CNR, being the largest and closest molecular structure to the SMBH, is  critical to understand fueling of both the SMBH and the nuclear star formation  \citep[e.g. see][ and references therein]{levin03,trani18}. The interplay between black hole feeding and star formation is therefore of paramount importance in the life-cycle of galaxies.

The CND of the Galactic Center (GC) is a molecular ring rotating with respect to the SMBH Sgr A*.
It has an uncertain gas mass between $10^{4}-10^{6}$ M$_{\odot}$ \citep{chris,great} and a size of $\sim$ 2 pc, within which are the ionized streamers called mini-spiral \citep[or Sgr A West;][]{roberts93,paumard04,zhao09} and nuclear star clusters \citep{figer99b}. Since Sgr A* is the closest SMBH, the CND in the GC therefore allows for the best spatial resolution to study the accretion processes in the local universe. The dynamical formation and destruction of the CND determines the lifetime of the cold gas accretion  surrounding Sgr A*. It has been well recognized that the tidal shear from the central potential is quickly rising within the central 10 pc \citep{gusten80,genzel85,jackson93,shukla04,chris}. The molecular clouds  cannot withstand tidal disruption if their densities are lower than the tidal limit, which is $10^{7}$ cm$^{-3}$ at a distance of 2 pc from Sgr A*.
Therefore, gas density measurements are crucial to understand whether the CND is stable and survives several orbital timescales ($\ge10^{5}$ yr). The gas densities estimated by the virial equilibrium for individual clumps range from 10$^{6}$ to 10$^{8}$ cm$^{-3}$ in earlier interferometric observations \citep{shukla04,chris,maria}. These virial densities are inconsistent with lower densities inferred from excitation analyses of spectral lines from single-dish observations \citep[10$^{4}$ to 10$^{6}$ cm$^{-3}$:][]{genzel85,marr93,bradford05,great,mills13b} and from dust observations \citep{white03}. Lower densities and masses are also reported in a recent ALMA observation \citep{tsuboi18}. Hence, these inconsistencies lead  to different conclusions regarding the lifetime of the CND.
However, the discrepancy between the density measurements might be the result of different observed scales in this complicated region \citep{great,mills13b,harada15}. The high resolution, sensitivity, and total power offered by ALMA allows us to revisit the measurements of physical properties of molecular clouds in the central parsecs of the GC.

The CND is not an isolated feature. The molecular gas surrounding the CND is resolved into multiple filaments with sizes $\sim$2 pc $\times$ 0.5 pc in the HCN(4-3) map (SubMillimeter Array; SMA) \citep{maria,liu12}.
Various molecular tracers also reveal a very clumpy distribution along the CND. The southern part of the CND is brighter than the northern part.
These streamers  originate from the ambient clouds 20-pc further out, and connect to the central 2 pc of the CND \citep{hsieh17}. The streamers are carrying gas  toward the CND and end up co-rotating with the CND. This kinematic analysis suggests that these streamers show a signature of ``infalling'' motions with progressively higher velocities as they approach the CND. The radial inward velocity is later indirectly confirmed by the magnetic  field configurations accounting for the observed dust polarization maps \citep{hsieh18}. These results suggest that the CND might have formed during tidal passages of nearby clouds. The broad linewidth of individual protrusions may also indicate the tidal stretching effect, i.e., the filaments are torn apart by the central gravity. This suggests that the streamers may be unstable and are dissolving at parsec-scale. These streamers contain numerous  compact dense cores with sizes of 0.05 pc in the previous ALMA  map \citep{hsieh19}. The next question is if these dense cores can survive the tidal disruption on their way of accretion onto the CND. Moreover, determining  whether these compact cores are prone to star formation near Sgr A* is also important to understand both the nature of the nuclear star clusters as well as the star formation process in a highly-dynamic environment around a SMBH.

\begin{figure}[ht!]
\epsscale{1}
\plotone{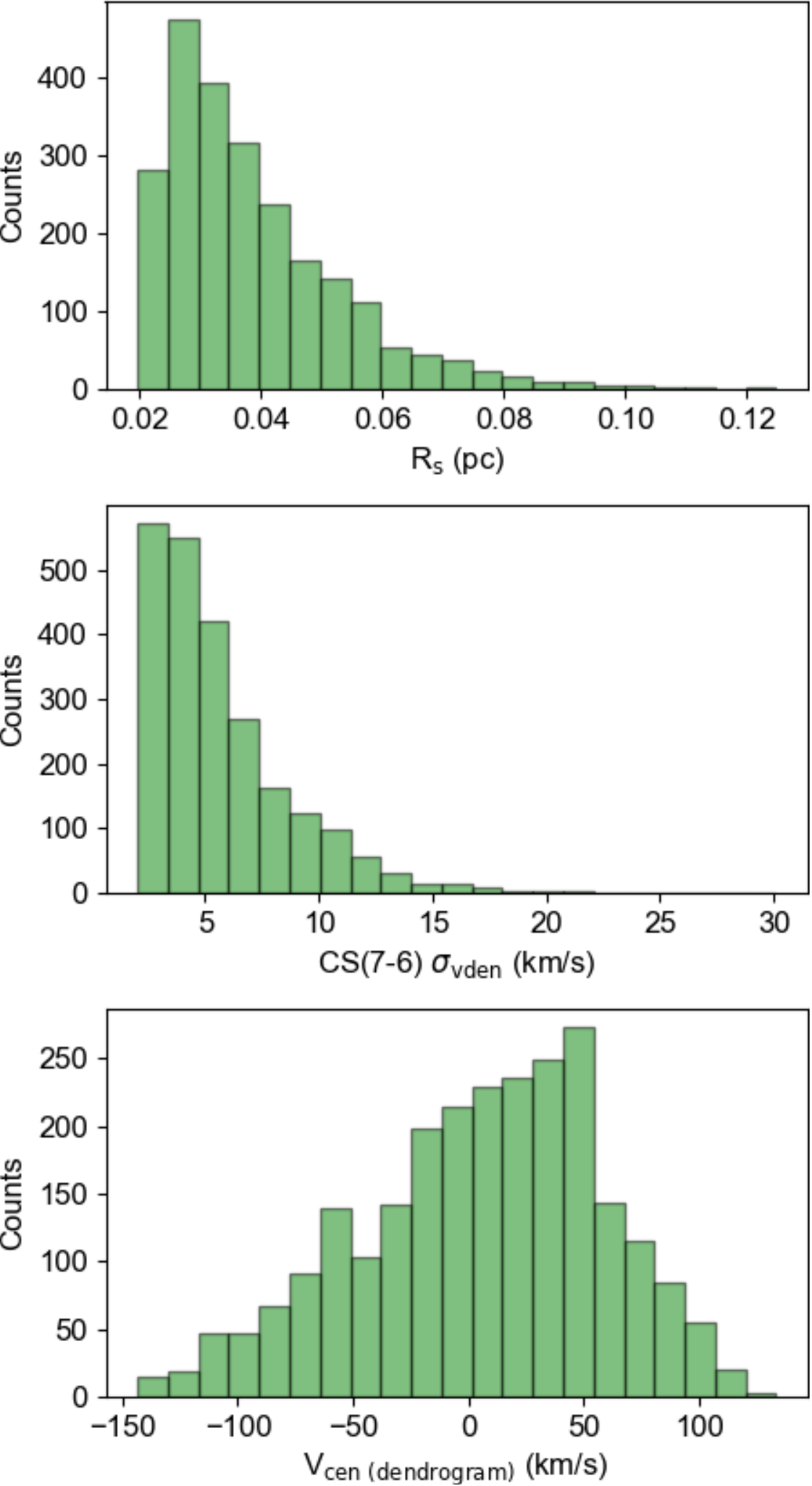}
\caption{Histograms of the effective spherical radius $R_{\rm s}$, rms linewidth ($\sigma_{\rm vden}$), and velocities of clumps identified with CS(7-6) line in astrodendro. The total number of identified clumps is 2379.
}
\label{fig-his-cld-dendro}
\end{figure}

\begin{figure}[ht!]
\epsscale{1}
\plotone{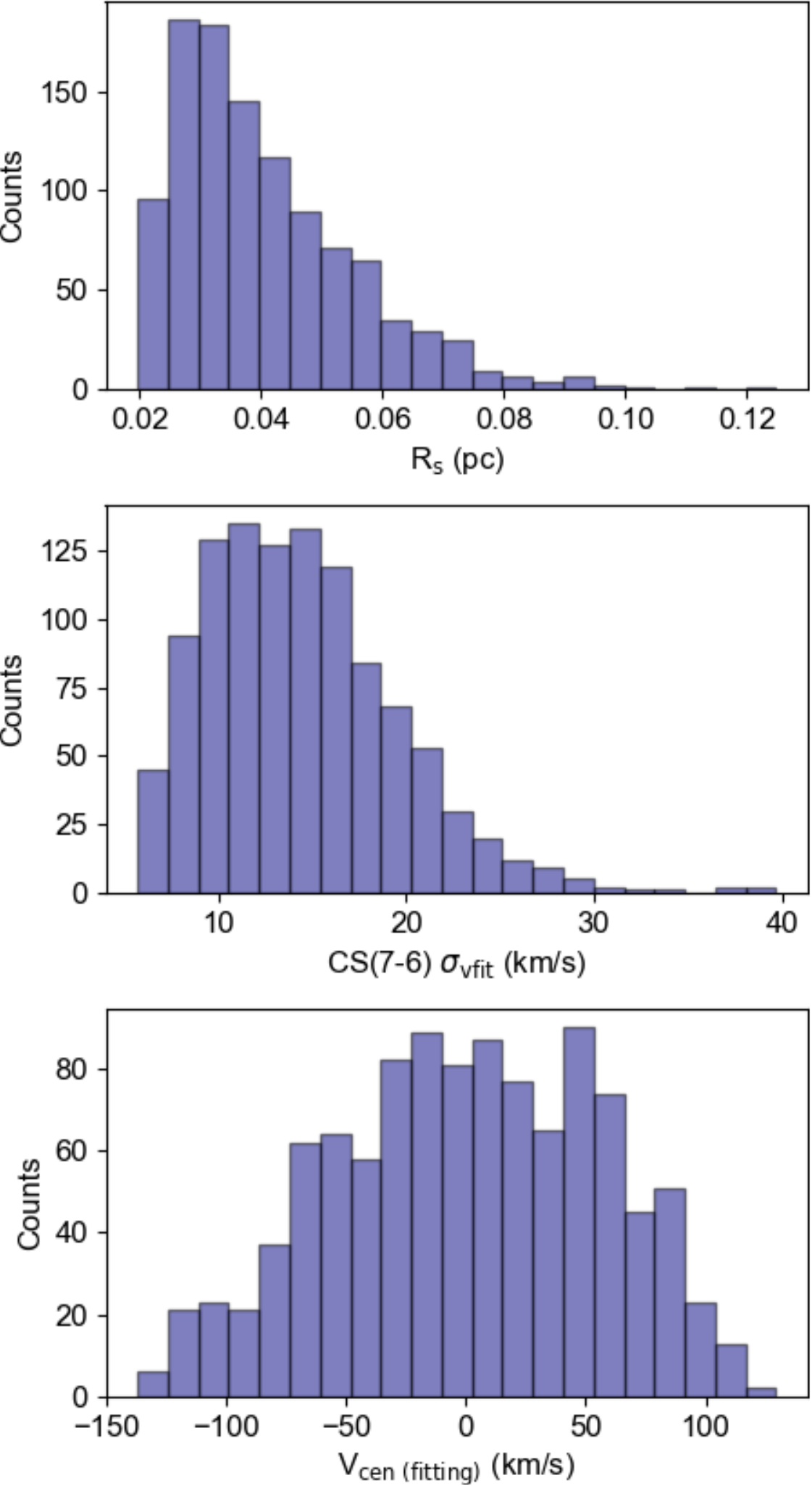}
\caption{Histograms of the effective spherical radius $R_{\rm s}$ derived from astrodendro, rms linewidth ($\sigma_{\rm vfit}$), and velocities of clumps identified with the CS(7-6) line with Gaussian fitting. The latter two parameters ($\sigma_{\rm vfit}$ and V$_{\rm cen~(fitting)}$) are derived from Gaussian fitting of spectra extracted at a leaf position.
The selection criterion for the fitting higher than $3\sigma$ is applied. This results in 1071 clumps, which is used for radex calculation. Gaussian fit to linewidth leads to larger linewidths than the astrodendro fit, see text.
}
\label{fig-his-cld}
\end{figure}

\subsection{Objectives of the ALMA Observations}
We have observed the CND and streamers with ALMA \citep{hsieh19}. Here we report a study of cloud clump properties within the inner 10 pc of the GC using millimeter and submillimeter lines of CS molecule. With these observations we can infer densities, kinetic temperatures and column densities for the clumps. With these data we can do a tidal stability analysis of the clouds do determine their likelihood of collapse. 
Utilizing the multiple CS transitions, we are able to measure the densities and sizes of the compact cores to derive a "surviving fraction" as a function of distance from Sgr A*. We can also probe the stability conditions of the compact cores in order to reveal the star formation conditions near a SMBH. Using the total power observation, we can determine the gas mass of the CND and streamers for the first time without previous assumptions to quantify the importance of the inflows and SMBH feeding. We have observed the CND and the streamers with the CS(7-6), (5-4), (4-3), (3-2), (2-1) lines in order to detect the dense gas accreting onto the CND. 
Multiple transitions with the same species are the key to determine the excitation conditions \citep[e.g.,][]{great,harada15}. The high-excitation lines minimize the self-absorption observed in lower transitions and more reliably sample the CND \citep{maria,tsuboi99,chris,wright01}.
CS is an ideal tracer to achieve the above listed objectives. CO, the second most abundant molecule is affected by foreground absorption even at J=4-3 \citep{great}. HCN, although commonly used to trace dense gas, is easily excited by  infrared radiation \citep{mills13b} or by electrons in some cases \citep{goldsmith17}. Hence, HCN is no longer a preferred density tracer. CS is  abundant and a more robust dense-gas tracer ($n_{\rm H_2}=10^{5-7}$ cm$^{-3}$) than  HCN and is more UV-resistant \citep{martin12}. We will use a radiative transfer model (RADEX) \citep{radex} to derive the physical parameters.  A detailed comparison of individual structures and kinematics will be presented in forthcoming papers.

\section{ALMA Observations and Data Reduction}\label{sect-obs}

\begin{center}
\begin{figure}[h!]
\epsscale{1.1}
\plotone{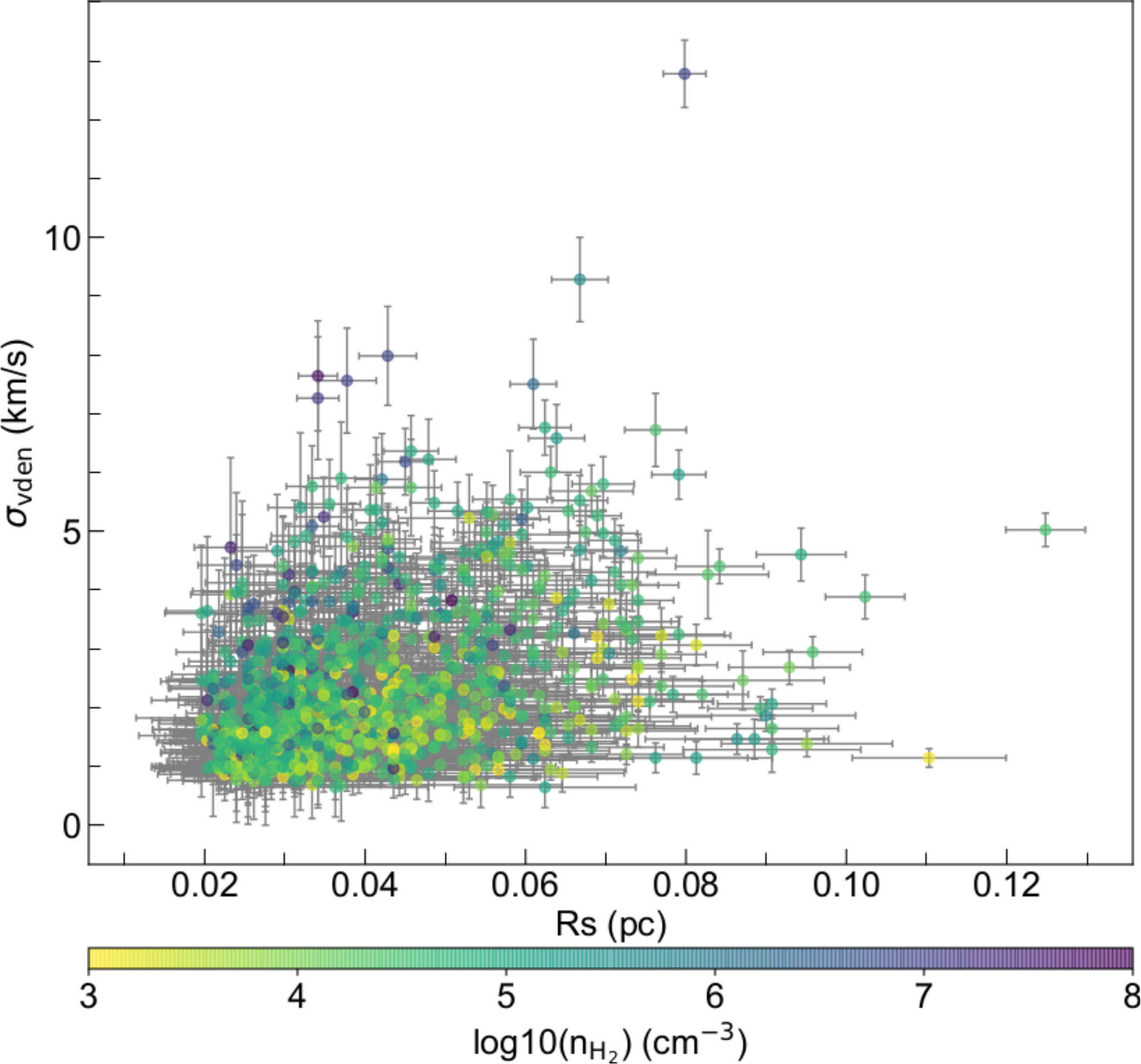}
\caption{Size-linewidth relation of the 1071 clumps are shown. $R_{\rm s}$ and $\sigma_{\rm vden}$ are shown in the x-axis and y-axis, respectively. Colors represent $n_{\rm H_2}$ derived in the later section.
}
\label{fig-dendro-m0}
\end{figure}
\end{center}

ALMA observations toward the central 3$\arcmin$ of the GC were carried out using the 12m array  and the total-power array (project code: 2017.1.00040.S; PI: Pei-Ying Hsieh). The total on source integration times are 7, 26, 60 hours for the 12-m array, the 7-m array, and single dish observations. We have observed the CS(7-6) ($f_{\rm rest}=$342.882 GHz), CS(5-4) ($f_{\rm rest}=$244.935 GHz), CS(4-3) ($f_{\rm rest}=$195.954 GHz), CS(3-2) ($f_{\rm rest}=$146.969 GHz), and CS(2-1) ($f_{\rm rest}=$97.980 GHz) lines (see Table \ref{tab:line}).
The default channel resolutions of the spectral windows are listed in Table \ref{tab:obs1}.
The observations consisted of a single field of a 68-pointing mosaic for CS(2-1) and 138 pointings CS(3-2). Two fields consisting of 251 and 300 pointings are performed to observe the CS(4-3) and CS(5-4) lines, respectively. Four fields consisting of CS(7-6) (600 pointing in total) are observed to cover the 3$\arcmin$ region. The details of the interferometric mosaics are listed in Table \ref{tab:obs1}, \ref{tab:obs2}, and \ref{tab:obs3}.

\begin{figure}[h!]
\epsscale{1.1}
\plotone{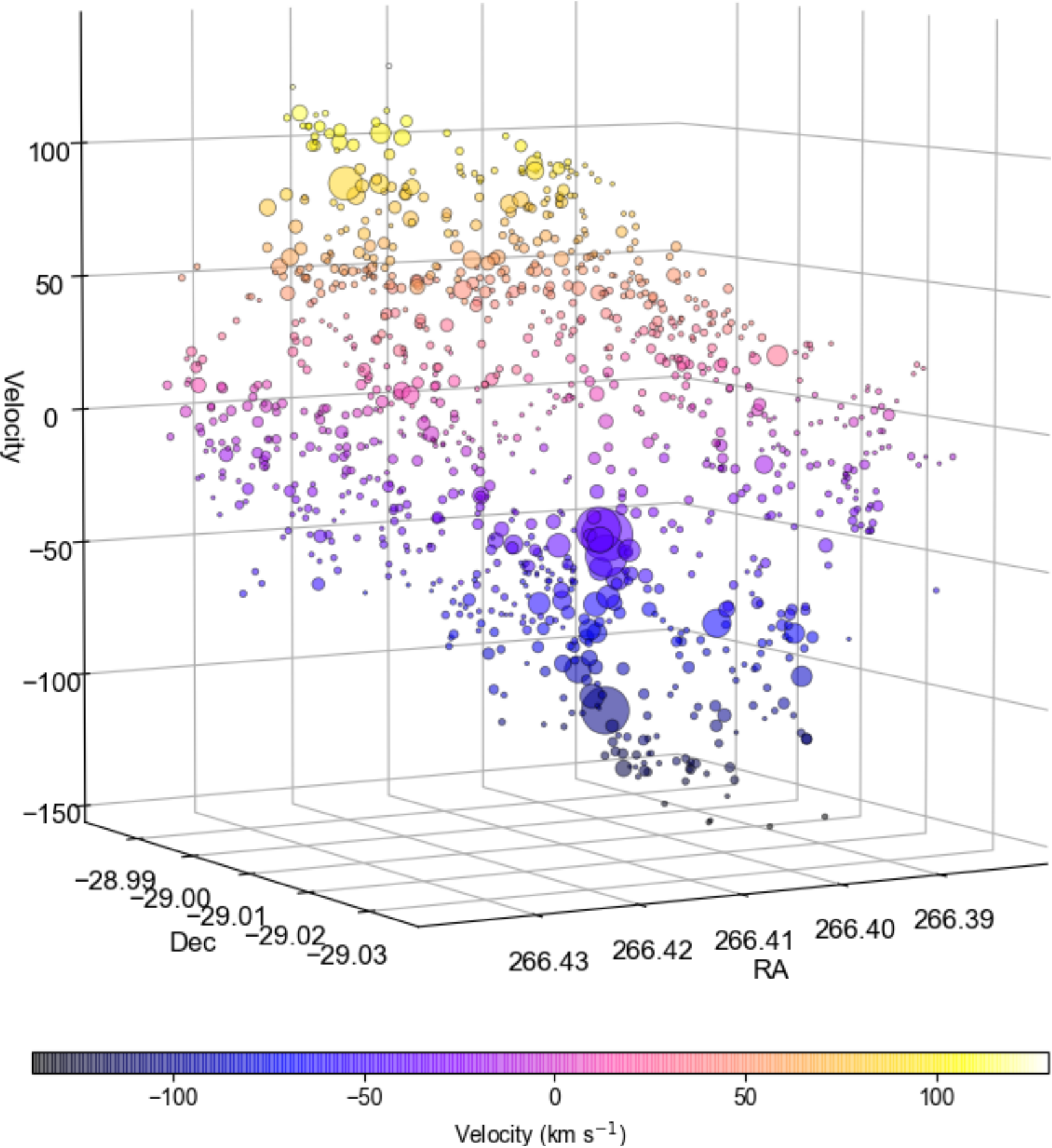} 
\caption{Position-position-velocity plot of the clumps. The color represents the centroid velocity and the sizes of circles are proportional to linewidth. All the properties are derived by Gaussian fitting.
}
\label{fig-3d}
\end{figure}

Calibration of the raw visibility was performed with the pipeline and manual reduction script for the cycle-5 data in Common Astronomy Software Applications (CASA v5.1.1-5). To recover the zero-spacing of the interferometric observations, we combined the 12 m, 7 m, and single dish data. The CASA (v5.4) tasks tclean and feather are used to deconvolve and merge the interferometric and single-dish maps. The automasking incorporated into tclean is applied to automatically mask regions during the cleaning process. The task Mstrasform is used to subtract the continuum emission. For observations of multiple fields, we clean and feather individual fields separately and combine individual fields to a single map. Weighted averages are applied in the overlapping area of adjacent fields.
We used the Briggs robust parameter of 0.5. The image cubes were made at a velocity resolution of 2 km s$^{-1}$ to enhance the signal to noise ratio.
The native resolution of the CS(2-1), CS(3-2), CS(4-3), CS(5-4), and CS(7-6) are 1$\arcsec$, 0.9$\arcsec$, 1.3$\arcsec$, 0.5$\arcsec$, and 0.8$\arcsec$, respectively. For comparisons of all lines, we smoothed the maps to 1.3$\arcsec$ ($=0.05$ pc), which is  the lowest resolution of the CS(4-3) map, for further analysis.  The noise level of the smoothed maps are 9.2, 10.8, 6.3, 32.4, 14.0 mJy beam$^{-1}$ for the CS(2-1), (3-2), (4-3), (5-4), and (7-6) maps, respectively.

\section{Results and Analysis}

\subsection{Integrated Intensity Maps}

\begin{figure}[bht!]
\epsscale{1}
\plotone{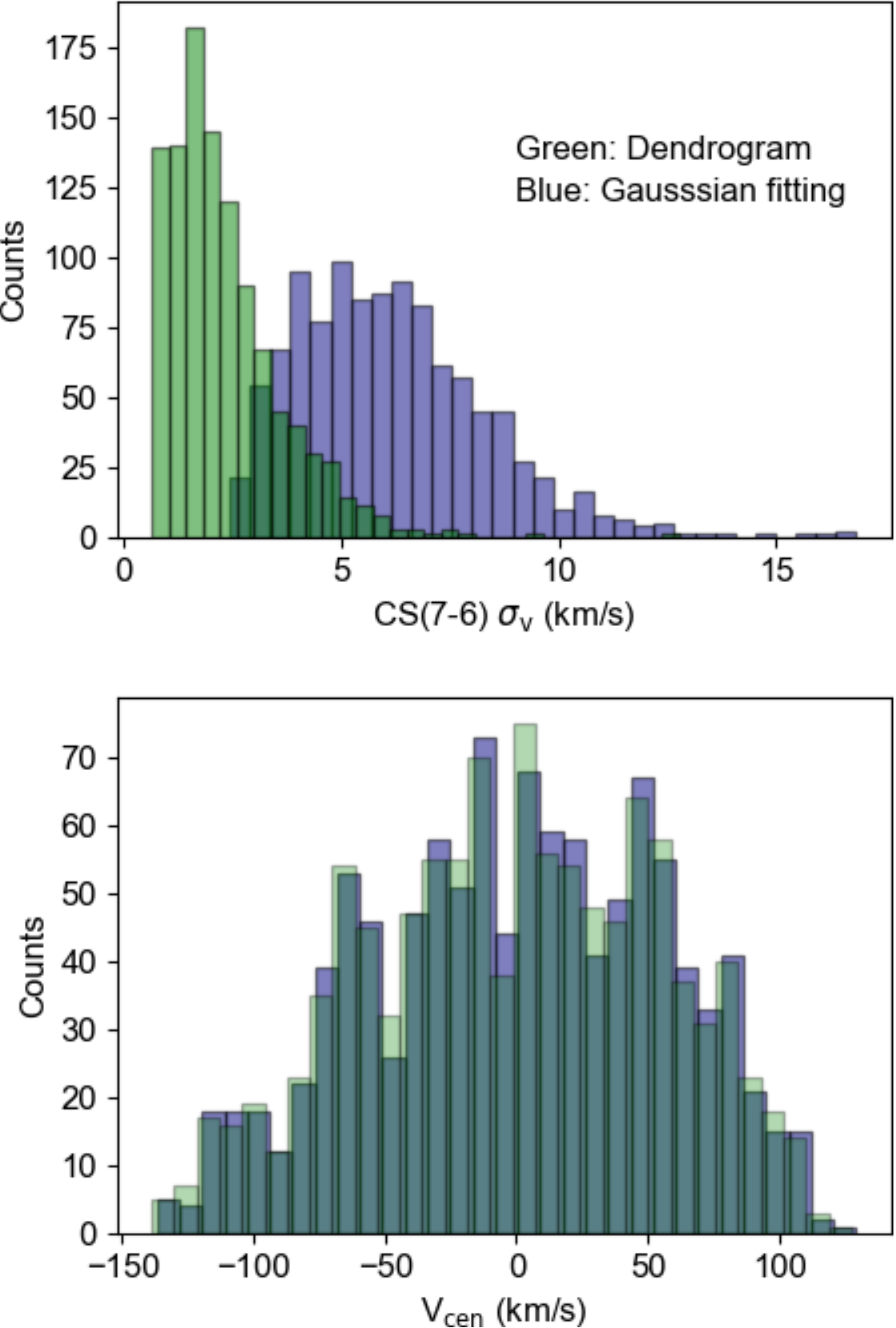} 
\caption{Histograms of the CS(7-6) $\sigma_{\rm v}$ and centroid velocity (V$_{\rm cen}$) derived from astrodendro (green) and  Gaussian fitting (blue). Dark green shows the overlapping area. The same clumps (number=1071) shown in Figure~\ref{fig-his-cld} are plotted here. Note that $R_{\rm s}$ is identical for the astrodendro and Gaussian approach as per definition. The linewidth of the Gaussian fits is larger due to underestimates by astrodendro, see discussion in text.}
\label{fig-his-cld-compare}
\end{figure}

\begin{figure}[bht!]
\epsscale{1}
\plotone{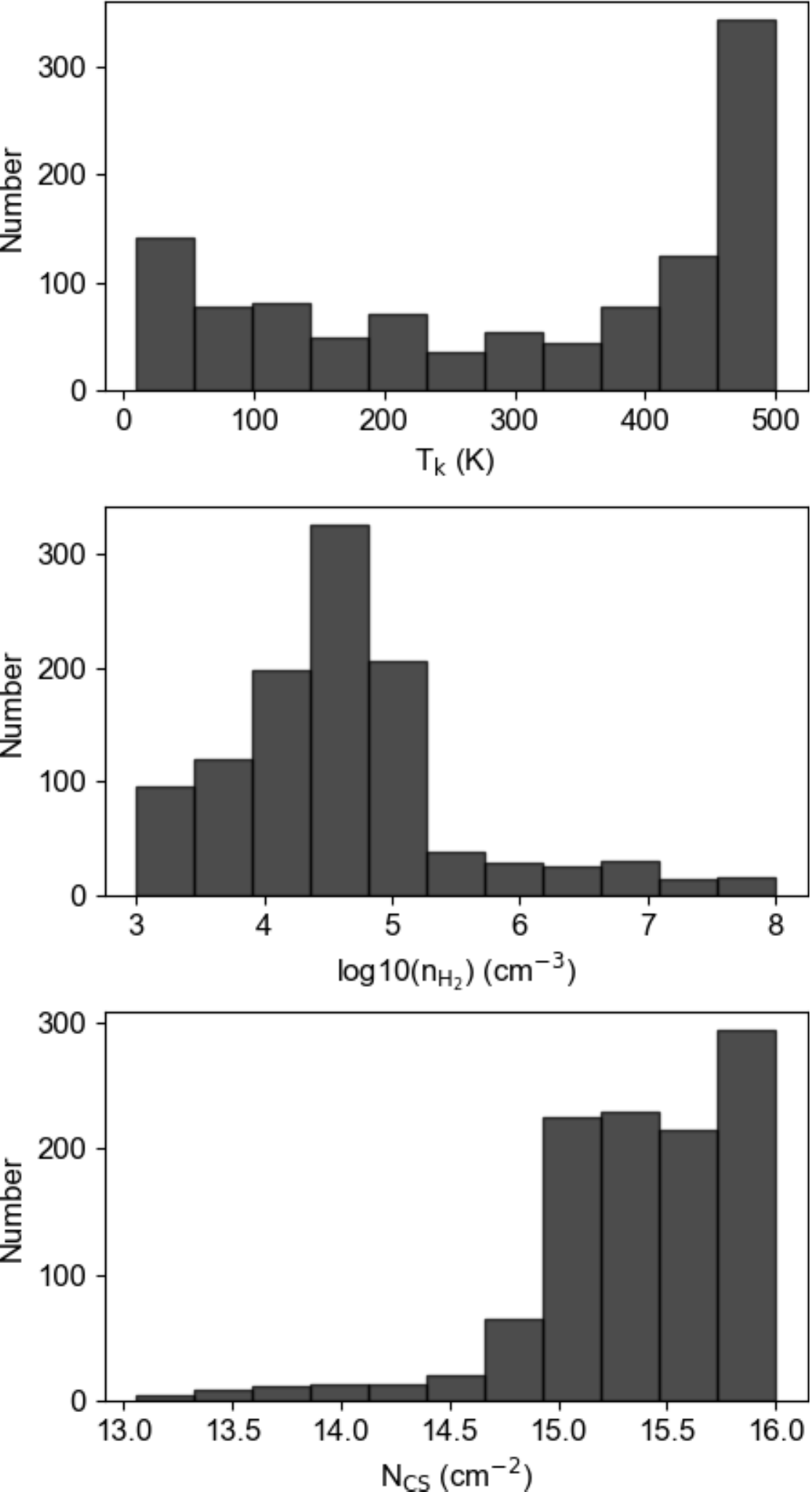}
\caption{Histograms of the derived $T_{\rm k}$, $n_{\rm H_2}$, and $N_{\rm CS}$ in Radex.}
\label{fig-his}
\end{figure}

\begin{figure*}[bht!]
\epsscale{1.1}
\plotone{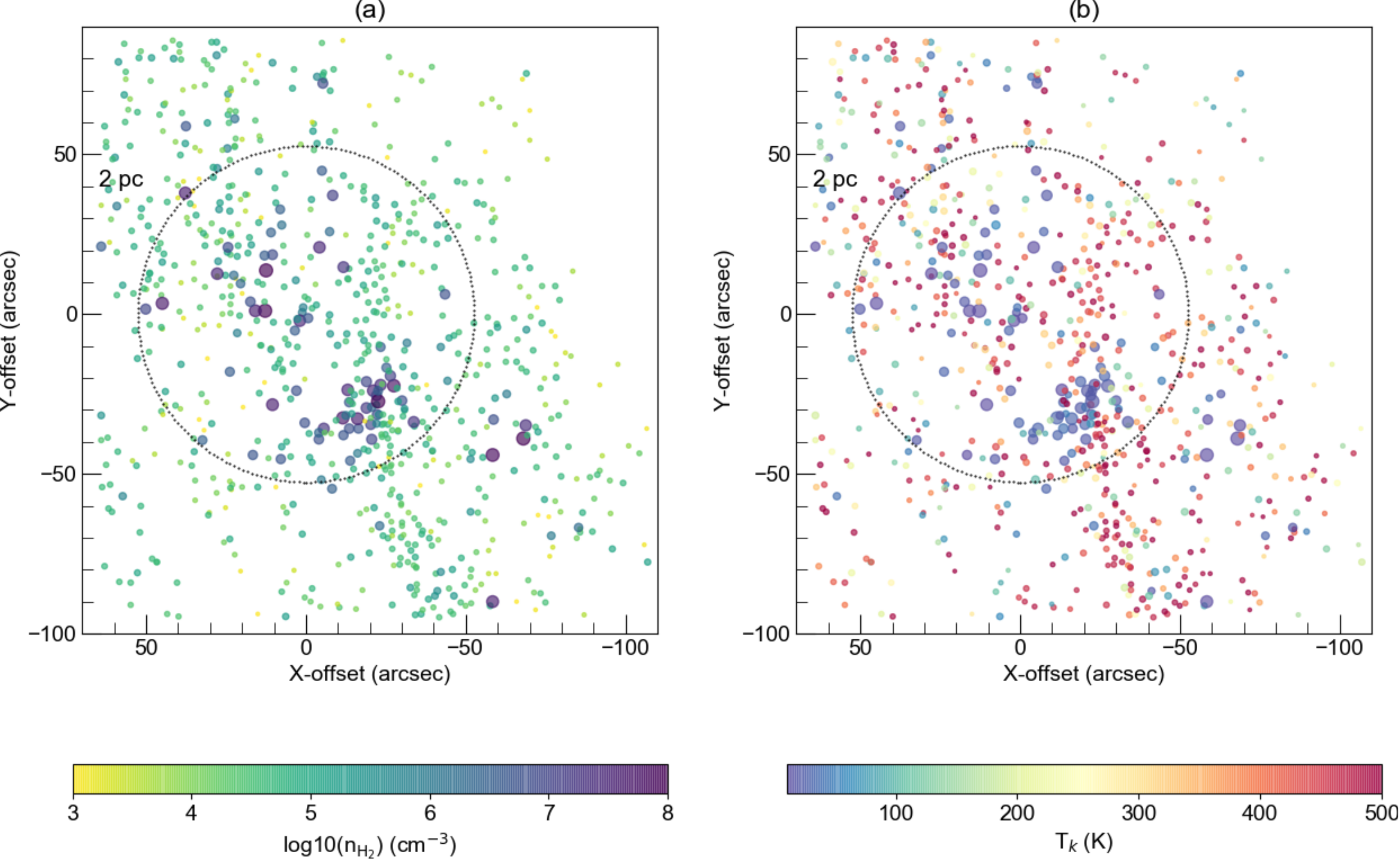}\vspace{0.4cm}
\caption{Clumps identified with {\it astrodendro} are shown in  RA-Decl. offsets, centered on Sgr A*. (a) Colors represent the density of the cloud, sizes of the circles are also proportional to density.  (b) Kinetic temperature in color. Sizes of circles are proportional to density, as in a). Radius of the dotted-circle is 2 pc.
}
\label{fig-proj-n-t}
\end{figure*}

\begin{figure*}[bht!]
\plotone{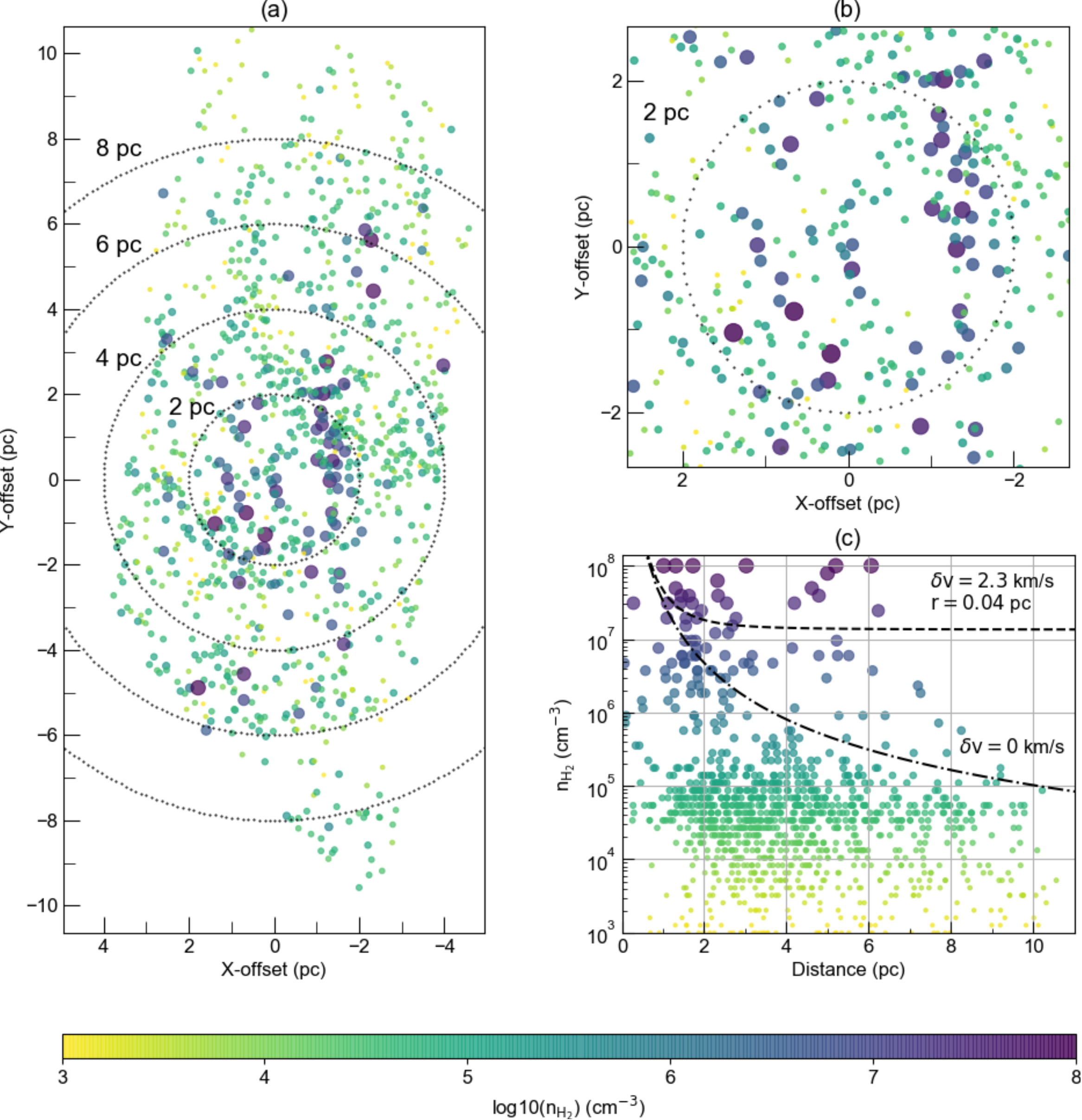}
\caption{Density of clumps as a function of the deprojected distance from Sgr A*. Color represents $n_{\rm H_2}$  in logarithmic scale  and the circle sizes are proportional to $n_{\rm H_2}$. Dotted circles mark the galactocentric distance of 2 to 8 pc. (a) Clump distribution in the deprojected space for the full range. (b) Same as (a), but for the central $\pm$2 pc. (c) $n_{\rm H_2}$ of clumps are shown as a function of the deprojected distance from Sgr A*.  The minimum density for tidal stability ($n_{\rm crit0}$) is labeled with the dot-dashed curve. The critical density ($n_{\rm crit}$) for gravitational collapse is labeled with the dashed curve for a cloud with a mean $R_\text{s}$ of 0.04 pc and a mean velocity dispersion of 2.3 km s$^{-1}$.
}
\label{fig-deproj-den}
\end{figure*}

\begin{figure*}[bht!]
\plotone{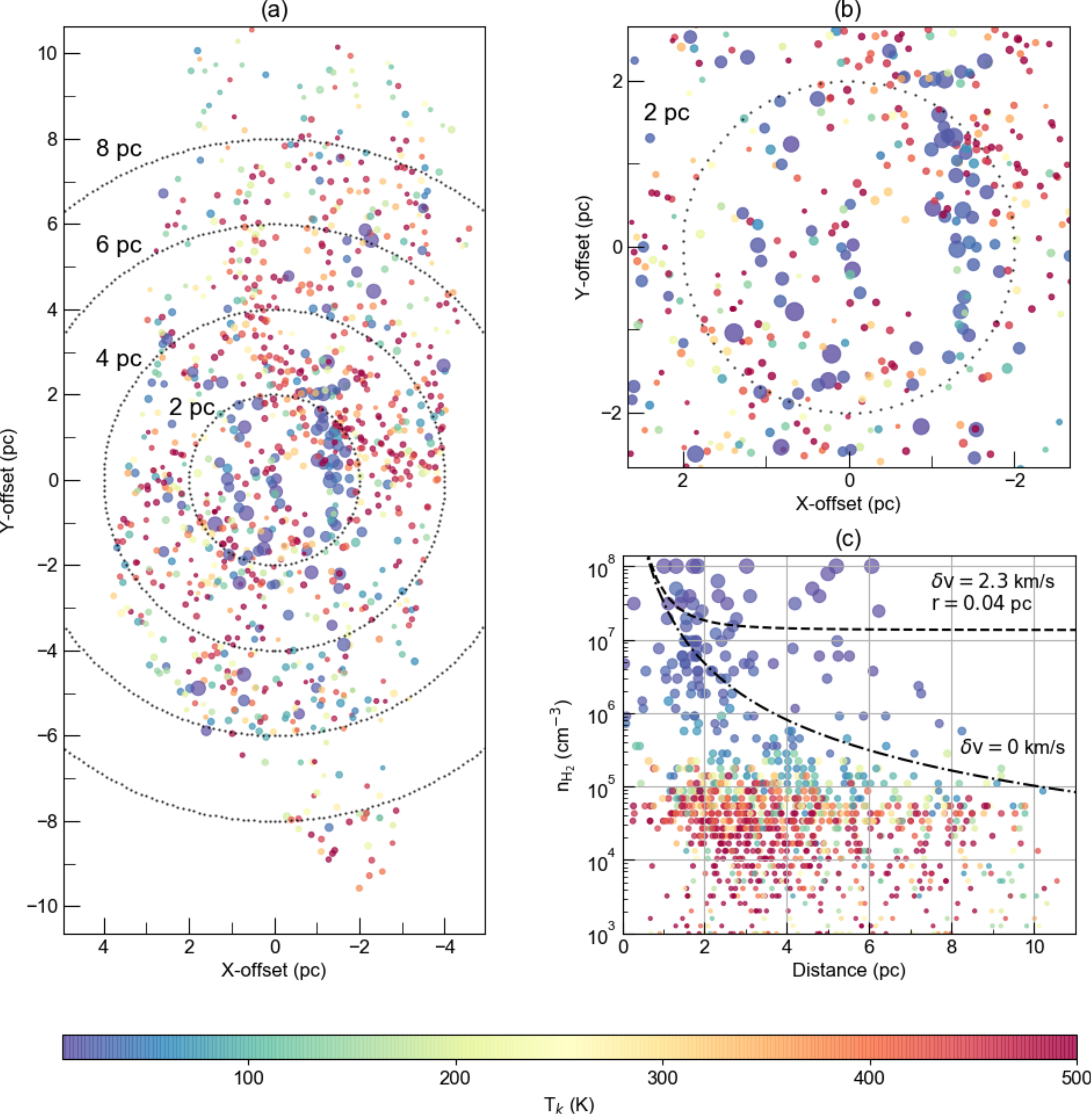}
\caption{As in Figure~\ref{fig-deproj-den}, but color represents $T_\text{k}$.
}
\label{fig-deproj-tk}
\end{figure*}

\begin{center}
\begin{figure*}[bht!]
\plotone{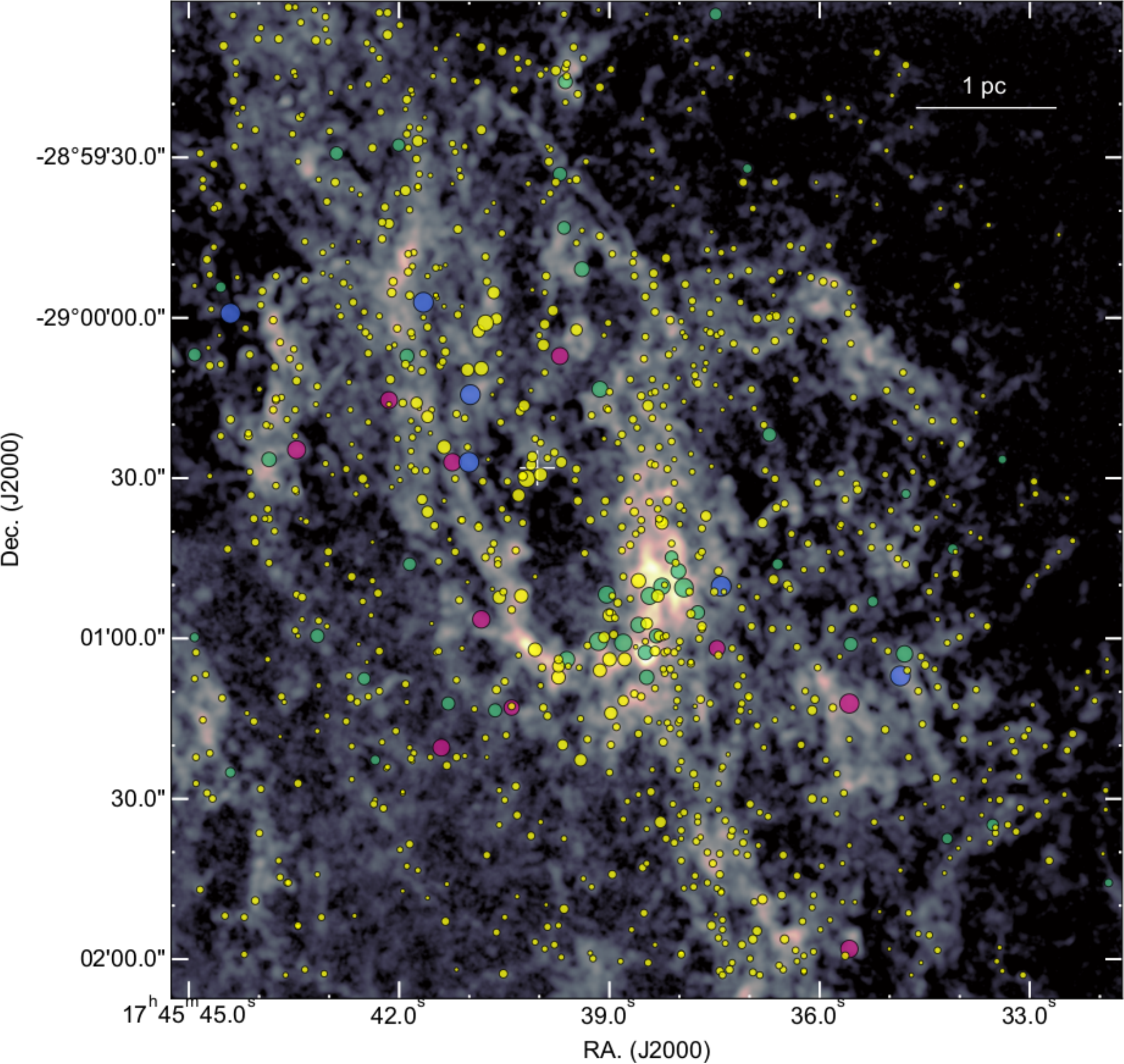}
\caption{Clumps overlaid  on the moment 0 map of the CS(7-6) line.
Yellow circles are clumps below $n_{\rm crit0}$ and will be dispersed by tidal force. Green circles are stable clumps within $n_{\rm crit0}$ and $n_{\rm crit}$ and will neither be dispersed by tidal force nor collapse by gravity. Red and blue circles are clumps above $n_{\rm crit}$ and able to collapse to form stars: the blue clumps correspond to the upper limit ($\leq 10^{8}$ cm$^{-3}$).  The circle sizes are proportional to $n_{\rm H_2}$.
}
\label{fig-11}
\end{figure*}
\end{center}

Figure~\ref{fig-m0} shows the integrated intensity maps of the CS(2-1), (3-2), (4-3), (5-4), and (7-6) lines, arising mostly from the CND and the surrounding streamers. A finding chart for the names of the features is presented in Figure~\ref{fig-m0}.
We follow the  nomenclature of previously identified features in \citet{chris,maria,martin12}.
The CND and its surrounding streamers are more clearly disentangled from the foreground and background materials with the higher excitation transitions, which are much more reliable for studying the structure of the CND. 
Comparing our data with the previous SMA maps \citep {maria,liu12,martin12},
the improved spatial resolution of our ALMA maps by a factor of about 4 in linear scale and by more than a factor of 10 in area resolves a great detail of sub-structure within the CND. Our maps show that the SMA 0.1 pc--0.25 pc sized clouds break into sub-parsec cores within the larger scale filaments. The CND no longer reveals itself as a complete ring but consists of multiple filaments wrapped around along the ``ring''.  The SW-lobe and western streamers are also resolved into multiple elongated filaments.  The molecular clumps within the central 0.5 pc of Sgr A* are resolved into multiple  filaments \citep{hsieh19}. The exquisite resolution in the ALMA maps displays a multitude of structures which lead to the impression that detailed dynamical features are now resolved and captured. Some areas show straight or clearly curved structures that appear to be tangential to rotation. We are likely  also witnessing scale-dependent structures that are clearly organized with finer and finer sub-structures. A network of filaments, sub-filaments, and even thinner fibers is resolved. These finer networks are also not random, but appear to connect to main filaments and streamers. The  CS(7-6) channel maps are shown in Appendix (Figure~\ref{fig-chan}). The detailed analysis of these structures and discussion of the channel maps will be presented in a forthcoming  paper. In this paper we will focus on the compact components in this region.

\subsection{CS Molecular Cloud Identification}

\subsubsection{Dendrogram}

We identify molecular clumps with the CS data cubes using the Python package {\it astrodendro}, which decomposes emission into a hierarchy of structures in the position-position-velocity (PPV) cubes \citep{rosolowsky08}. 
The dendrogram decomposes the molecular structures into branches and leaves. Branches are structures that can be split into sub-structures called leaves, which have no further resolved substructure. The largest continuous structures surrounding the local maxima were categorized as trunks (i.e., no parent structures). We adopt parameters so that {\it astrodendro} identified local maxima in the cube above a threshold of $5~{\sigma }_{\mathrm{rms}}$, and at least $2.5~{\sigma }_{\mathrm{rms}}$ above the merging level of the branches with adjacent structures. In addition to the intensity threshold for decomposition, we also specify the minimum number of pixels that a structure should span to be the area of one synthesized beam, and remove the clumps with linewidths smaller than 2 km s$^{-1}$.

The primary goal is to decompose spectroscopic components superimposed along the line of sight in this complicated region. Since the high excitation line is expected to be less biased by the foreground emission, to cleanly identify the ``leaves'' associated with the CND and streamers, we identify clumps with the CS(7-6) data.
The numbers of the leaves (hereafter clumps) identified by {\it astrodendro} are 2379.
The basic properties of the identified clumps are calculated by {\it astrodendro}. The root-mean-square (rms) size of the major and minor axes of clumps ($\sigma_{\rm maj}$, $\sigma_{\rm min}$), and the rms linewidth ($\sigma_{\rm vden}$) are determined by the moment calculations.
The calculations of moments are based on pixels assuming that the cloud is continuous and bordered by an isosurface of a certain brightness temperature.  The position and velocity of the clumps are determined by the intensity-weighted centroids of pixels within a cloud.
The sizes of the clumps ($\sigma_{\rm r}$) are defined by the geometric mean of $\sigma_{\rm maj}$ and $\sigma_{\rm min}$ ($\sigma_{\rm r}$=$\sqrt{\sigma_{\rm maj}~\sigma_{\rm min}}$).  Multiplying $\sigma_{\rm r}$ with a coefficient 1.91 converts the rms size to the effective spherical radius of the cloud ($R_{\rm s}$) \citep{solomon87,bertoldi92}. The histograms of these basic properties are shown in Figure~\ref{fig-his-cld-dendro}.

We use a bootstrap approach to calculate the uncertainties in the derived properties \citep{rosolowsky06}. For each cloud, we generate a trial cloud and measure its properties ($\sigma_{\rm maj}$, $\sigma_{\rm min}$, $\sigma_{\rm v}$) by random sampling pixels (allowed repetition) within the boundaries of the cloud in position and velocity space. We repeat this process 500 times and use the median absolute deviation to estimate an rms uncertainty for each property. The uncertainty is scaled by the square root of the number of pixels of the synthesized beam to account for the oversampling rate.

\subsubsection{Gaussian Fitting}
To measure the total flux of every transition in a homogeneous area and position, using the  centroid positions and velocities of clumps identified in the CS(7-6) cube, we extract spectra from the cube averaged over the $\sigma_{\rm maj}$ and $\sigma_{\rm min}$, and fit a single Gaussian profile to every CS transition. Single Gaussian is in general a good approximation of the clumps. The fitted intensity peaks and  linewidths are then used to calculate the flux for analysis of radiative transfer modeling (next section). We have selected clumps where all transitions simultaneously show a fitting result higher than 3 $\sigma$. However, we find that including CS(2-1) leads to fewer clumps passing the filter of 3 $\sigma$ (748 clumps). To improve the statistical results, we exclude CS(2-1). We also exclude the clumps with a difference of linewidths larger than 5 km s$^{-1}$, which is a 2.5 $\sigma$ limit of the velocity resolution.  The resulting number of clumps is 1071, which accounts for 54\% of the CS flux of the 2379 clumps. The histograms of the effective spherical radius (R$_{\rm s}=$$\sigma_{\rm r}\times$1.91), the rms linewidth of the CS(7-6) ($\sigma_{\rm vfit}$), and the  central velocity derived from Gaussian fitting  are shown in Figure~\ref{fig-his-cld}. The size-linewdith relation of these 1071 clumps is shown in Figure~\ref{fig-dendro-m0}. We also present the cloud properties in PPV space in Figure~\ref{fig-3d}.  It is clearly seen that the clumps in the southern/northern part of the CND (blue/red shifted velocities), and within the streamers have broader linewidths than the surrounding gas.

Comparisons of the velocity dispersion and centroid velocity given by {\it astrodendro} and our Gaussian fitting are shown in Figure~\ref{fig-his-cld-compare}. The centroid velocities are consistent with each other. Note that in {\it astrodendro}, the $\sigma_{\rm r}$ of a leaf is defined by the high isocontour levels (bijection method). This leads to an underestimated
$\sigma_{\rm v}$ because the spectral profile is truncated by the isosurface boundary \citep{rosolowsky06}. This effect can be seen in Figure~\ref{fig-his-cld-compare}. The $\sigma_{\rm v}$ given by the {\it astrodendro} is smaller than our Gaussian fitting (at the same position). Also note that we do not correct for the resolution effect, this is because a simple deconvolution of the clump size estimated by the bijection method may be underestimated, as discussed in \citet{rosolowsky06}.

\subsection{Statistical Equilibrium and Non-LTE Radiative Transfer}

We use the radiative transfer code Radex \citep{radex} to derive the physical properties of the molecular gas in the CND and its surrounding.
Radex performs the statistical equilibrium calculations and uses the escape probability formalism \citep{goldreich} to solve the statistical equilibrium equations to model the observed line intensities for given physical conditions.
The statistical equilibrium calculations account for the opacity effects as well as subthermal excitation.

The general solution of the radiative transfer equation for a homogeneous medium is,
\begin{equation}\label{bb}
I_{\nu} = B_{\nu}(T_{\rm ex})(1-e^{-\tau_{\nu}})+I_{\nu}^{\rm bg}e^{-\tau_{\nu}},
\end{equation}
where $I_{\nu}$ is the specific intensity and $I_{\nu}^{\rm bg}$ is the background emission at line frequency $\nu$, in units of erg s$^{-1}$ cm$^{-2}$ Hz$^{-1}$ sr$^{-1}$. $T_{\rm ex}$ and $\tau_{\nu}$ are the excitation temperature and the optical depth, respectively. $B_{\nu}(T)$ is the blackbody radiation field at temperature $T$.
The intensity of the line subtracting the background emission ($I_{\nu}-I_{\nu}^{bg}$) is calculated  for each molecule with varying molecular column density, H$_{2}$ density of the main collision partner, and kinetic temperature.
Note that the Rayleigh-Jeans approximation is generally not valid for molecular line emission at submm wavelengths of thermal emission.  We thus calculate  $I_{\nu}$  and do not use the Rayleigh-Jeans temperature calculated by the code. In this paper, we use the uniform sphere geometry, a single collision partner (H$_{2}$), and a background temperature of 2.73 K. The collisional rate and spectroscopic data of the CS molecule were taken from Leiden Atomic and Molecular Database (LAMDA) \citep{lamda,lique}.

\begin{figure*}[bht!]
\epsscale{1.1}
\plotone{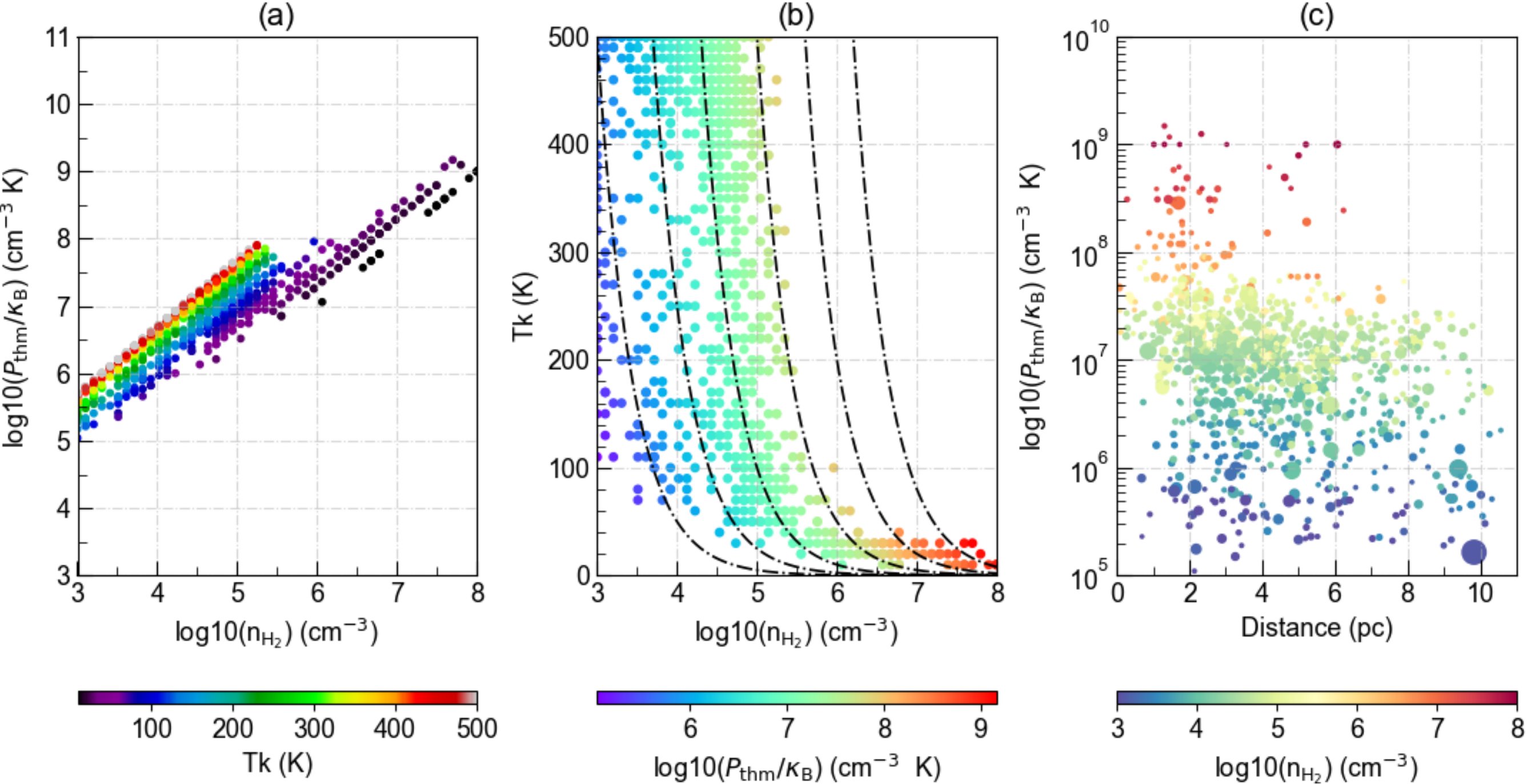}
\caption{(a) Thermal pressure of clumps $P_{\rm thm}/k_{\rm B}$ (cm$^{-3}$ K) \emph{vs.} $n_{\rm H_2}$. Color denotes $T_{\rm k}$. The high density clumps ($\ge 10^{6}$ cm$^{-3}$) do not have high temperature components as shown in the lower density clumps.
(b) $T_{\rm k}$ \emph{vs.} $n_{\rm H_2}$. Colors indicates $P_{\rm thm}/k_{\rm B}$. Isobaric curves of $P_{\rm thm}/k_{\rm B}$ = $5\times10^{5}$, $2.5\times10^{6}$,  $1\times10^{7}$, $5\times10^{7}$, $2\times10^{8}$, and $8\times10^{8}$ cm$^{-3}$ K are labeled with the dot-dashed curves.
(c) $P_{\rm thm}/k_{\rm B}$ \emph{vs.} the distance from Sgr A* with color indicating $n_{\rm H_2}$. The circle  size is proportional to  $R_{\rm s}$, showing that higher density clumps have smaller size.
}
\label{fig-thermal}
\end{figure*}

We construct a 50$\times$50$\times$50 grid of parameter space for kinetic temperature ($T_{\rm K}$), molecular hydrogen density ($n_{\rm H_2}$), and column density of CS molecule ($N_{\rm CS}$).
\begin{enumerate}
\item $T_{\rm K}$ is from 10 K to 500 K.
\item $n_{\rm H_2}$ is from 10$^{3}$ cm$^{-3}$ to 10$^{8}$ cm$^{-3}$.
\item $N_{\rm CS}$ is from 10$^{12}$ cm$^{-2}$ to 10$^{16}$ cm$^{-2}$.
\end{enumerate}

We then derive the best-fit $T_{\rm K}$, $n_{\rm H_2}$ , and $N_{\rm CS}$ by comparison with the velocity integrated intensities (line flux) $\int I_{\nu}dv=1.0645I_{\nu}\Delta v$ , where $\Delta v$ is the full width half maximum (FWHM) of the line, and the factor of 1.0645 is a correction from the adopted square line profile in Radex to Gaussian profile. $I_{\nu}$ is derived by Equation~\ref{bb}. The model integrated intensity is conserved at each specific condition. The same amount of radiation is spread over the spectral profile, and therefore $I_{\nu}$ (also called line intensity peak in observation), depends on linewidth because the opacity changes. The opacity tends to be lower with broader linewidths because the same amount of radiation is spread over wider velocities. The variation of opacity as a function of linewidth is not linear, especially in higher density conditions. We therefore generate the Radex models by incrementing $\Delta v$ with 5 km s$^{-1}$ intervals from 5 km s$^{-1}$ to 40 km s$^{-1}$, and fit the data at corresponding intervals. The comparison of the models and the observed four transitions (excluding CS(2-1)) is performed with the $\chi^2$-statistics for each set of parameters. We find the best fit models by minimizing $\chi^2$ value.


The brightness unit measured in the radio map is often defined in Jy per beam (Jy/beam) which is the same dimension as specific intensity (erg s$^{-1}$ cm$^{-2}$ Hz$^{-1}$ beam$^{-1}$) but averaged over the Gaussian beam (synthesized beam). Therefore, to compare with the model, we need  Jy beam$^{-1}$ to be a more physical unit of per steradian. We divide by the solid angle of the beam in steradian ($\Omega_{\rm beam}$).
\begin{equation}
 \langle I_{\nu} \rangle (\text{Jy sr}^{-1}) = \frac{I_{\nu}(\text{Jy beam}^{-1})}{\Omega_{\rm beam}},
\end{equation}
\begin{equation}
\Omega_{\rm beam} = \frac{\pi\theta_{b1}\theta_{b2}}{4\ln2}\left(\frac{\pi}{180^{\circ}\times3600}\right)^{2},
\end{equation}
where $\theta_{b1}$ and $\theta_{b1}$ are the gaussian major and minor axis of the synthesized beam in units of arcsec.

However, several uncertainties arise for unknown source sizes and non-uniform structures. With an insufficient resolution, it is not possible to avoid these uncertainties. For unresolved sources (compact sources or point sources), the values in Jy/beam are equal to the flux density (Jy) because the total fluxes are covered within one beam. Therefore, to calculate the specific intensity from flux density, we should divide the measured value ($I_{\nu,\rm map}$) by the solid angle of source sizes ($\Omega_{\rm source}$). Sources smaller than half of the synthesized beam are almost indistinguishable and the accuracy is less than 10\%. Therefore, for such sources  a lower limit is given from dividing by the beam size, because the source sizes are unknown. If the sources are resolved, the assumption is that sources have a uniform intensity distribution. To optimize the comparison with the model, we introduce a scaling factor in the fitting to correct for the uncertainty of intensity from the above reasons. This is similar to a beam filling factor but not exactly the same because we do not use a temperature scale. The best fitted scaling factors are between 0.05 to 2.5.

Histograms of the derived $T_{\rm k}$, $n_{\rm H_2}$, and $N_{\rm CS}$ from the Radex analysis are shown in Figure~\ref{fig-his}. The majority of the clumps have  $T_{\rm k} \ge 400$ K, and show a smooth distribution between 50 K to 400 K. Gas with $n_{\rm H_2} \le$ 10$^{5}$ cm$^{-3}$ dominates the region.  Overall, 78\% of the compact clumps in this region have  $T_{\rm k} \ge 50$ K and $n_{\rm H_2} \le$ 10$^{5}$ cm$^{-3}$.
High temperature and low density environments in the CND were reported in previous studies using single-dish data.  \citet{bradford05}, derived warm 200-300 K and moderate densities 5-7$\times10^{4}$ with CO(7-6). \citet{oka11} reported $T_\text{k}=63$ K and $n_{\rm H_2}=10^{4}$ cm$^{-3}$ with CO(1-0) and CO(3-2).  \citet{great}  constrained the temperatures of the southwest lobe of the CND to $T_\text{k}=200$, $n_{\rm H_2}=10^{4.5}$ cm$^{-3}$ for a low excitation component, and $T_\text{k}=500$, $n_{\rm H_2}=10^{5.2}$ cm$^{-3}$ for a high excitation component with the nine CO transitions.
Using the HCN(3-2), (4-3), and (8-7) lines, \citet{mills13b} showed that the southern CND has $T_\text{k}=94-171$ K and $n_{\rm H_2}=10^{5.9-6.5}$ cm$^{-3}$, and the northern CND has $T_\text{k}=270$ K and $n_{\rm H_2}=10^{5.6}$ cm$^{-3}$. \citet{harada15} derived the $n_{\rm H_2}=7.26\times 10^{5}$ cm$^{-3}$ and $N_{\rm CS}=1.07\times 10^{15}$ cm$^{-2}$ with the CS(2-1), (6-5), (7-6), (8-7) lines in the southern lobe of the CND.
It appears that with the CS and HCN lines, higher densities are determined than CO.
With the high resolution ALMA data, we further reveal the existence of cooler and denser components with $T_\text{k}\leq 50$, $n_{\rm H_2}=10^{6-8}$ cm$^{-3}$ at sub-pc scale. In conclusion, the CND consists of a mixture of warm ($T_{\rm k}\ge 50-500$ K, n$_{\rm H_2}$=$10^{3-5}$ cm$^{-3}$) and cold gas ($T_{\rm k}\le 50$ K, n$_{\rm H_2}$=$10^{6-8}$ cm$^{-3}$) \citep{nh3}.

\begin{figure*}[bht!]
\epsscale{1.1}
\plotone{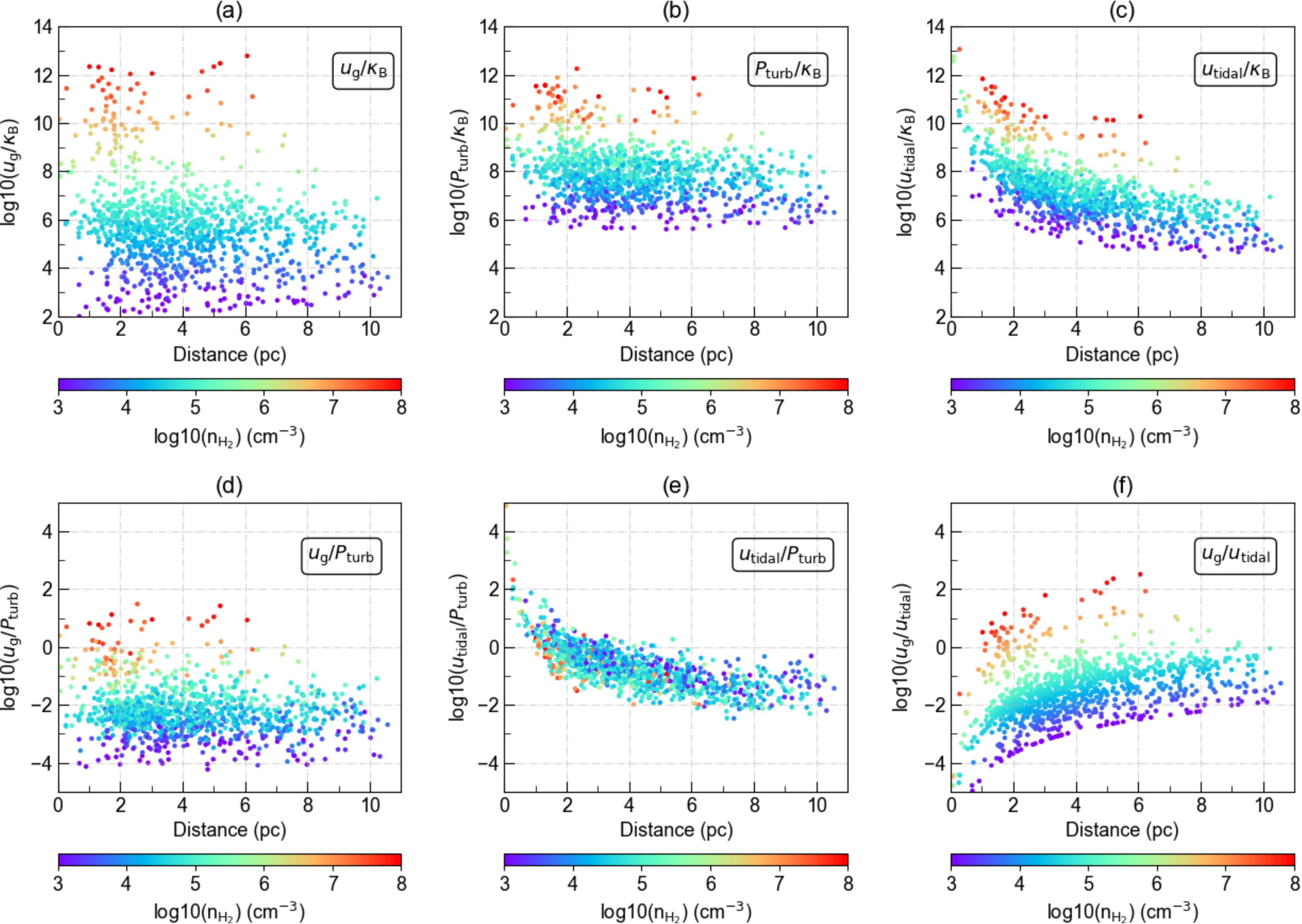}
\caption{(a) Gravitational energy density $u_{\rm g}$, (b) turbulence 
pressure $P_{\rm turb}$, and (c) energy density of tidal force ($u_{\rm tidal}$) as a function of the galactocentric distance from Sgr A*.  Ratios among two energy components are displayed in the lower panels. 
}
\label{fig-energy}
\end{figure*}

\begin{figure*}[bht!]
\epsscale{1.1}
\plotone{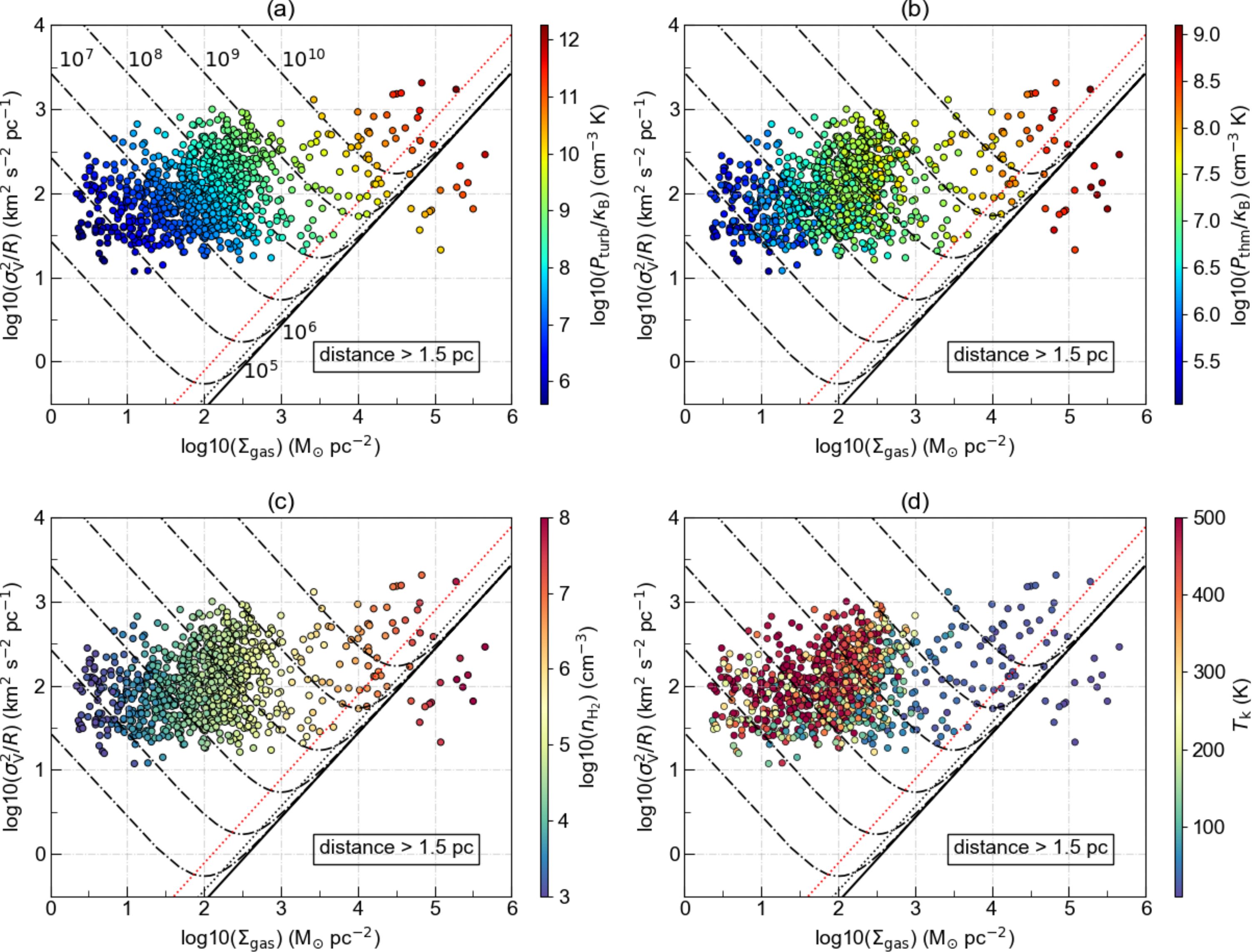}
\caption{(a) Distribution of the clumps in the $\sigma_v^2/R - \Sigma_\text{gas}$ plane. 
The identified clumps (at distance from Sgr A* $>$ 1.5 pc) are overlaid colored by their turbulent pressure. The dot-dashed curves draw Equation (16) with varying external pressure ($P_{\rm ext}/k_{\rm B}=10^{5-9}$ cm$^{-3}$ K), ignoring magnetic fields and tidal force. The solid straight line is the solution with $P_\text{ext}=0$.
The black and red dotted lines (corresponding to critical configurations) show the location of critical mass of clumps determined from the assumptions of constant density (Equations 2 and 3 in \citealp{field11}) and centrally concentrated internal density structure (Equations 4 and 5 in \citealp{field11}), respectively. Clumps above the dot-dashed lines at corresponding $P_{\rm ext}$ and $\sigma_{\rm v}^{2}/R$ are gravitationally unstable.
(b), (c), (d) Same as (a) but colored by thermal pressure $P_{\rm thm}$, $n_{\rm H_2}$, $T_{\rm k}$, respectively.
}
\label{fig-pext}
\end{figure*}


\section{Discussion}

In Figure~\ref{fig-proj-n-t} we show the spatial distribution of the $n_{\rm H_2}$ and $T_{\rm k}$ derived from the Radex model on the projected sky. The ALMA high resolution data reveal a broad distribution of densities and temperatures at sub-pc scale associated with the CND and the surrounding streamers.
In addition, these features contain numerous  compact dense cores with sizes of $\le$0.05 pc in the ALMA maps. A key question is whether these dense cores can survive the tidal disruption near Sgr A*. With the high resolution ALMA map, we will derive the cloud properties at sub-pc scale.

\subsection{Stability of Clumps as a Function of Distance from Sgr A*}

To compare the gas densities with the tidal limit -- the minimum density for tidal stability, as a function of distance from Sgr A*, we deproject  the distances of clumps with respect to Sgr A*. The inclination and position angle used for deprojection are 70$^{\circ}$ and 30$^{\circ}$, respectively.  The inclination of 70$^{\circ}$ is the median between the 65$^{\circ}$ to 80$^{\circ}$ of the streamers and the CND \citep{martin12,hsieh17}.
The spatial distributions of densities and temperatures of clumps in the deprojected plane  are shown in Figure~\ref{fig-deproj-den}(a), (b) and figure~\ref{fig-deproj-tk}(a), (b), respectively.  In the deprojected map, the radii of 2, 4, 6, and 8 pc are shown as dotted circles. The CND is located mostly between 2 to 3 pc, while the streamers are located between 2 to 8 pc. This deprojection assumes that the structures are aligned in the same plane \citep[cf.][]{hsieh19}.
We find that there is no clear trend or correlation of the $n_{\rm H_2}$ and $T_{\rm k}$ as a function of the distance from Sgr A*. Our results appear consistent with the previous studies by \citet{chris}, who  derived core densities at a scale of 0.25 pc and  reported that there is no correlation between $n_{\rm H_2}$ and the distance from Sgr A*.

To find the critical density for disruption by tidal shear due to Sgr A* and the nuclear star clusters, we use the model from \citet{vollmer00}, for the enclosed mass
\begin{equation}
M_{\rm total} = M_{\rm BH}+1.6 \times 10^{6} r_{\rm pc}^{1.25} M_{\odot},
\end{equation}
where $M_{\rm BH}=4\times 10^{6}$ M$_{\odot}$ is the mass of Sgr A* \citep{ghez05} and $r_{\rm pc}$ is the radial distance from Sgr A* in units of pc.
The tidal force at the position $x$ from the cloud center located at $r$ is given by
\begin{equation}
\begin{aligned}
T(x)&=-x\frac{d}{dr}\left(\frac{GM_{\rm total}}{r^{2}}\right)\\
&=x\frac{GM_{\rm BH}}{(\rm 1pc)^{3}}(2r_{\rm pc}^{-3}+0.3r_{\rm pc}^{-1.75})\\
&=x\mathcal{T}_{0}(r),
\end{aligned}
\label{eq-tidal}
\end{equation}
where $\mathcal{T}_0 = 2GM_{\rm BH} r^{-3}(1 + 0.15 r_{\rm pc}^{1.25})$.
For a cloud with mass  $M$ and radius $R$ ($=R_{\rm s}$), tidal disruption occurs if
\begin{equation}
\frac{GM}{R^2} < R\mathcal{T}_{0}(r).
\end{equation}

Assuming the cloud is uniform, its mass is given by
\begin{equation}
M=1.36m_{\rm H_{2}}n_{\rm H_{2}}\left(\frac{4\pi R^3}{3}\right),
\end{equation}
where 1.36 is a correction for He and other elements, and $m_{\rm H_2}$ is the molecular hydrogen mass. Assuming that the cloud has uniform density, the  critical density ($n_{\rm crit0}$) of molecular hydrogen against tidal force is then given by
\begin{equation}
n_{\rm crit0} = 2.8\times10^{7}\left(r_{\rm pc}^{-3}+0.15r_{\rm pc}^{-1.75}\right)~{\rm cm^{-3}},
\label{eq-crit}
\end{equation}
which is the tidal limit. In Figures~\ref{fig-deproj-den}(c) and \ref{fig-deproj-tk}(c), we plot  $n_{\rm crit0}$ as a dot-dashed line. Clumps above this line are tidally stable. A vast majority of the high temperature clumps (with $T_{\rm k}\ge50$ K) have densities  smaller than  $n_{\rm crit0}$ in the central 10 pc of the GC. The cooler components (with $T_{\rm k}\leq 50$ K) have in general higher density than the $n_{\rm crit0}$. We also find that within the central 1 pc, none of the clumps are above $n_{\rm crit0}$.

\subsection{Effects of Turbulence}

The above analysis does not take into account the effects of internal and external pressures on the cloud stability. In the presence of the external  force $\mathbf{F}_{\rm ext}$, the scalar virial theorem in the equilibrium state for an unmagnetized gas reads
\begin{equation}
0 = 2U+W-P_{\rm ext}\int_{S} \mathbf{x}\cdot d\mathbf{A}+\int_{V} \rho\mathbf{x}\cdot \mathbf{F_{\rm ext}}d^{3}x,
\label{eq-virial1}
\end{equation}
where $U$ is the internal energy, $W = -3GM^{2}/(5R)$ is the gravitational potential energy, $P_{\rm ext}$ is the pressure of an external medium, and $F_{\rm ext} = T(x)$ is the (external) tidal force. Assuming that the turbulent pressure dominates the thermal pressure inside the cloud, Equation~\eqref{eq-virial1} becomes
\begin{equation}
0 = 3M \sigma_{\rm v}^{2}-\frac{3GM^{2}}{5R}-4\pi R^{3} P_{\rm ext} + \frac{3}{5}\mathcal{T}_{0}MR^{2},
\label{eq-virial}
\end{equation}
where $M$ is the cloud mass, $R=R_\text{s}$, $\sigma_{\rm v}$ is the velocity dispersion, and $P_{\rm ext}$ is the external pressure. Self-gravity and external pressure tend to confine the cloud against the internal turbulence and tidal force tending to disperse the cloud.

For a time being, we assume $P_{\rm ext}=0$ (we will discuss the effect of $P_{\rm ext}$ in the later section). Then, Equation~\eqref{eq-virial} gives
\begin{equation}
\begin{aligned}
n_{\rm crit} = 2.8&\times10^{7}\left(r_{\rm pc}^{-3}+0.15r_{\rm pc}^{-1.75}\right)~{\rm cm^{-3}}\\
&+4.1\times10^{7}\left(\frac{\sigma_{\rm v}}{{10~\rm{km~s}^{-1}}}\right)^{2}
\left(\frac{R}{0.1 \rm{pc}}\right)^{-2}~\rm{cm}^{-3},
\end{aligned}
\label{eq-crit-virial}
\end{equation}
which is the critical density including tidal force and turbulence. This is identical to Equation~(\ref{eq-crit}) if $\sigma_v=0$. Using the results of {\it astrodendro}, we overlay $n_{\rm crit}$ (dashed line) using the mean values of $\sigma_{\rm v}$=2.3 km s$^{-1}$ and $R_\text{s}$=0.04 pc in Figures~\ref{fig-deproj-den}(c) and  \ref{fig-deproj-tk}(c).
Here we plot all the clumps. Figure~\ref{fig-d-error} in Appendix plots in the density-distance plane the clumps with uncertainties of densities less than 1 order of magnitude.
The critical densities $n_{\rm crit}$ and $n_{\rm crit0}$ define 3 different regimes.
\begin{enumerate}
\item Clumps below $n_{\rm crit0}$ are tidally disrupted  and will disperse.
\item The clumps that  fall in between $n_{\rm crit0}$ and $n_{\rm crit}$ are tidally stable and at the  same time supported by turbulence against gravitational collapse .  Collapse is still not happening and with growing dispersion these clumps could also disperse.
\item The clumps above $n_{\rm crit}$  are tidally  stable and have densities large enough that they can collapse. The free-fall time is roughly a few thousand years (see Figure~\ref{fig-4panel} in Appendix).
\end{enumerate}
Note that the curve derived by the mean values of $\sigma_\text{v}$ and $R_\text{s}$ only presents an overall threshold of $n_{\rm crit}$. The classification of clumps in the 3 different regimes based on the individual value of the clump is presented in Figure~\ref{fig-11}. The fractions of gas mass and number in each type are presented  in Table~\ref{tab:mass}.  At the distance from Sgr A* $>$ 1.5 pc, the internal energy (turbulence) is more important than the tidal force, and supports clumps against gravity. However, within the central 1.5 pc, tidal force overrides again  and the threshold densities for gravitational collapse quickly rises to $\ge10^{8}$ cm$^{-1}$.

Clumps below $n_{\rm crit0}$ account for $16${\raisebox{0.3ex}{\tiny$\substack{+28 \\ -7.5}$}} \% of the total mass in our samples ($2.5${\raisebox{0.5ex}{\tiny$\substack{+1.0 \\ -0.4}$}}$\times10^{4}$  M$_{\odot}$). Note that this mass is a lower limit. 94\% of the clumps (by numbers) belong to this regime, suggesting that most of the clumps are tidally unstable.
Among the clumps we identified, nearly $84${\raisebox{0.3ex}{\tiny$\substack{+16 \\ -37}$}} \% of the total gas mass is tidally stable ($28${\raisebox{0.3ex}{\tiny$\substack{+24 \\ -12}$}} \% of the total mass is gravitationally stable and $56${\raisebox{0.3ex}{\tiny$\substack{+22 \\ -28}$}} \% of the total mass is gravitationally unstable).
The 16 seemingly gravitationally unstable clumps are listed in  Table~\ref{tab:sfcloud}. The 36.2 GHz and 44.1 GHz class I methanol masers were observed in the CND \citep{yusef08,sjo10,pih11}. The class I methanol masers are collisionally pumped and are associated with shocked gas or outflows, which could be a signature of early phase of star formation \citep[cf.][]{yusef08,yusef15}. However, there is no  850$\micron$ continuum emission detected in association with these masers, and no HCN gas associated with corresponding velocities \citep[e.g.][]{liu13}. These masers were suggested to be produced by the shock during cloud-cloud interactions \citep{sjo10}. It has not yet been concluded that a star is about to form in the CND or the star formation is quenched. If the star formation is quenched in the CND, additional forces such as internal magnetic pressure/tension, rotational shear, or cloud rotation \citep{vollmer00} may  exist to support the clouds against collapse.

\subsection{Thermal and Non-thermal Properties of the Clouds}

We calculate the thermal pressure ($P_{\rm thm}$) of the clumps with the derived gas density and kinetic temperature as
\begin{equation}
P_{\rm thm}/k_{\rm B}=n_{\rm H_2}T_{k},
\end{equation}
where $k_{\rm B}$ is the Boltzmann constant. The relations between $P_{\rm thm}-n_{\rm H_2}$ and $T_{\rm k}-n_{\rm H_2}$ are shown in Figure~\ref{fig-thermal}. 
The $P_{\rm thm}-n_{\rm H_2}$ relation resembles a typical phase diagram for diffuse ISM and has a break at $n_{\rm H_2,crit} = 3\times10^{5}$ cm$^{-3}$, such that high density clumps with $n_{\rm H_2} > n_{\rm H_2,crit}$ are cold with $T_{\rm k} < 50$ K, while low density clumps with $n_{\rm H_2} < n_{\rm H_2,crit}$  are relatively warm with $T_{\rm k} > 50$ K.  Thermal pressure of low density clumps varies by an order of magnitude for fixed $n_{\rm H_2}$.
Figure~\ref{fig-thermal}(b) shows that the temperature monotonically decreases as the density increases, implying that pressure varies a lot from cloud to cloud. Figure~\ref{fig-thermal}(c) hints that the pressure of clumps is overall systematically growing towards Sgr A*, i.e., the highest pressure 
clumps are found nearest to Sgr A*, while lower-pressure
clumps are at larger distances.

Interestingly, we find that there is no high density cloud with high temperature. If high density clumps are able to form stars, the lack of   high temperature gas may hint an maximum mass of newly-forming stars because the Jeans mass is proportional to $T^{3/2}$. It is possible that the CS transitions we observed are not sensitive to high temperature and high density gas, and we need to observe the mid-infrared high-J lines to probe these missing populations. However, since the cooling rate is proportional to the square of gas density, and if the dust-gas coupling is better toward high density clouds, the cooling can be efficient \citep{krumholz11}.

We further calculate the energy densities of turbulence, tidal force, and gravity to compare their relative importance acting on the clumps. 
The gravitational energy density is given by
\begin{equation}
u_{\rm g}=\frac{9}{20\pi}\frac{GM^2}{R^{4}}.
\end{equation}
The turbulence energy density is
\begin{equation}
P_{\rm turb}=\frac{3}{2}\rho\sigma_{\rm v}^{2},
\end{equation}
where $\rho$ is the mass density of gas and $\sigma_{\rm v}$ is from {\it astrodendro}.  The energy density of the tidal force is given by 
\begin{equation}
u_{\rm tidal}=\frac{9}{20\pi}\frac{\mathcal{T}_{0}M}{R}.
\end{equation}
Figure~\ref{fig-energy} plots the energy densities and their ratios. In contrast to $u_{\rm tidal}$, which is a decreasing function of $r$, we find that   $u_{\rm g}$ and $P_{\rm turb}$ do not change much with $r$, although there is a growing 
number of clumps with an increased level of turbulence
closer to Sgr A*. This is similar to the trend seen in the thermal pressure. 

Figure~\ref{fig-energy} shows that for most clumps, 
$P_{\rm turb} > u_{\rm tidal} > u_{\rm g}$ at $r>$ 2 pc except for the densest clumps.  In what follows, we ignore $u_{\rm tidal}$ in comparison with $P_{\rm turb}$. We note that although the clumps are warmer in the GC, turbulence is still more important than the thermal energy. The ratio $u_{\rm tidal}$/$P_{\rm turb}$ depends on the galactocentric distance, but independent of the cloud density. Therefore, the relative importance of each term is dependent on the mass of the black hole and turbulence. We note  that the tidal force can fully override turbulence for a black hole mass 100 times larger than Sgr A*. The physical condition and distribution of energy in external galaxies can be largely different from Sgr A*.

\subsection{Can External Pressure Stabilize Clouds?}

To study the role of the external pressure ($P_{\rm ext}$) on confining molecular clouds, we compare our data with the model presented by \citet{field11}.
In their model, a non-magnetized virial theorem immersed in a uniform external pressure is applied. The pressure-bounded virial equilibrium (PVE) requires 
\begin{equation}\label{eq:field1}
\frac{\sigma_{\rm v}^{2}}{R}=\frac{1}{3}  \left(\pi\Gamma G\Sigma_{\rm gas}+\frac{4P_{\rm ext}}{\Sigma_{\rm gas}} \right),
\end{equation}
where $\Gamma$ is a geometry factor (0.6 for a sphere of constant density and 0.73 for an isothermal sphere of critical mass), $\sigma_{\rm v}^{2}/R$ is the scaling coefficient, and $\Sigma_{\rm gas}= M/ (\pi R^2)$ is the gas surface density. Note that Equation~(\ref{eq:field1}) is analogous to Equation \eqref{eq-virial} in the absence of the tidal force.  For a given external pressure and kinetic energy, the critical mass (maximum mass) of a cloud in equilibrium is obtained by
\begin{equation}\label{eq:field2}
\begin{aligned}
\frac{\sigma_{\rm v}^{2}}{R_{\rm c}}&=\frac{1}{3}  \left(\pi\Gamma G\Sigma_{\rm c}+\frac{4P_{\rm ext}}{\Sigma_{\rm c}} \right)\\
&=(6.0~\text{or}~6.4) \times10^{-12}\left(\frac{P_{\rm ext}}{k} \right) ~{\rm cm~s^{-2}}
\end{aligned}
\end{equation}
where $R_{\rm c}$ and $\Sigma_{\rm c}$ are the critical radius and critical surface gas density, respectively.
The factor $6.0$ and $6.4$ in Equation \eqref{eq:field2} are  for centrally concentrated internal density and uniform density, respectively. The cloud mass lower than the critical mass will be gravitationally stable. We refer to Equations (2) to (5) of \citet{field11} for the derivation of Equation (17). In this model, the effect of tidal force is not considered. 
We still can make a comparison with our data as we find that the tidal force is less important in the virial equilibrium  at the distance $\ge$ 1.5 pc (see  Figure~\ref{fig-deproj-den}).

Figure~\ref{fig-pext} plots the clumps (at $r\ge 1.5$ pc) in the ${\sigma}_{\rm v}^{2}/R-\Sigma_{\rm gas}$ plane. The dot-dashed lines draw Equation~(\ref{eq:field1}) in the range of $P_{\rm ext}/k_{\rm B}=10^{5-10}$ cm$^{-3}$ K. The solid line corresponds to $P_{\rm ext}=0$. In Figure~\ref{fig-pext}(a), (b), the non-thermal pressure of the clumps 
is $10-10^4$ times higher than the thermal pressure.
If the clumps exist in a range of different pressure environments, the clumps leftward to the critical lines could be stable if the required $P_{\rm ext}$ exists to confine them. Some clumps located in the region between the critical line and the $P_{\rm ext}=0$ line are in unstable configurations such that small perturbations would drive them to gravitational collapse.  Clumps located to the right of the $P_{\rm ext}=0$ line are not in virial equilibrium, and would collapse if there are no other supporting forces such as magnetic forces, which are not considered in the present work.
Figure~\ref{fig-pext} shows that a majority of the clumps we identified are in PVE if the external pressure is in the range of $P_{\rm ext}/k_{\rm B}=10^{6-9}$ cm$^{-3}$ K. The molecular clouds in the central molecular zone (CMZ) are also proposed to be in pressure equilibrium \citep{miyazaki00,oka01}. 
In the GC, the external pressure of ambient X-ray amouning to $P_{\rm ext}/k_{\rm B}=$(1-2)$\times10^{6}$ cm$^{-3}$ K \citep{ponti19} is found and would be able to confine the clouds with densities of $10^{3-4}$ cm$^{-3}$, even if they have densities lower than $n_{\rm crit0}$.
Since these low density clumps are the majority population of the observed morphology, confinement via external pressure ensures a relatively steady morphology.

The external pressure required for equilibrium in the CND is 2-4 orders of magnitude higher than the typical molecular clouds in galactic disks \citep[$P_{\rm ext}=10^{4-5}$ cm$^{-3}$ K,][]{bertoldi92,belloche11}. If the CND is pressure-bounded, a much higher external pressure is required. External magnetic fields with a strength of 1 mG can provide an external pressure of $\sim 10^{8}$ cm$^{-3}$ K. In addition, the hot gas from the mini-spiral also provides external pressure of $\sim 10^{8}$ cm$^{-3}$ K inside the CND. However, it is unclear whether these two sources provide isotropic and homogeneous confinement since (1) the magnetic field strength is subject to scale of structures and we are still unaware whether the magnetic field plays an important role on core-scale in the CND, and (2) the mini-spiral only provides confinement of the molecular clouds with narrow velocity and space.  High resolution magnetic field observations made by ALMA are important to further study the CND-clouds.

\section{Summary} 

We have presented ALMA and TP array observations of 5 CS rotational lines ($J_\text{u}=$7,5,4,3,2) in the central 10 pc of the GC. With a spatial resolution of 1.3$\arcsec=0.05$ pc, we are able to probe the gravitational stability of molecular clumps within the filamentary structures. Using the {\it astrodendro}, we identify 2379 clumps. Among these clumps, 1071  clumps are able to derive the n$_{\rm H_2}$ and $T_{\rm k}$ using Radex.

\begin{itemize}
\item The compact clouds are a mixture of warm ($T_{\rm k}\ge 50-500$ K, n$_{\rm H_2}$=$10^{3-5}$ cm$^{-3}$) and cold gas ($T_{\rm k}\le 50$ K, n$_{\rm H_2}$=$10^{6-8}$ cm$^{-3}$). 
\item 78\% of the compact clumps (fraction of number) in this region have $T_{\rm k} \ge 50$ K and $n_{\rm H_2} \le$ 10$^{5}$ cm$^{-3}$.
\item The total gas mass of these 1071 clumps is $2.5${\raisebox{0.5ex}{\tiny$\substack{+1.0 \\ -0.4}$}}$\times10^{4}$  M$_{\odot}$. 1009 clumps are tidally unstable and account for $16${\raisebox{0.3ex}{\tiny$\substack{+28 \\ -7.5}$}} \% of the total gas mass. However, we find that although the tidal force may affect the cloud morphology of the lower density gas, a part of them can be confined by external pressure from X-ray emission.
\item $28${\raisebox{0.3ex}{\tiny$\substack{+24 \\ -12}$}} \% of the total gas mass is tidally stable, but will not collapse to form stars.
\end{itemize}

Based on the unmagnetized virial theorem including tidal force, we study the behavior of the molecular gas and the possibility of star formation in the CND.

\begin{itemize}
\item  At a distance $\ge$ 1.5 pc from Sgr A*, turbulence dominates the internal energy and sets the nearly constant threshold densities of $3\times 10^{7}$ cm$^{-3}$, which inhibits the clumps from collapsing to form stars.
\item  $56${\raisebox{0.3ex}{\tiny$\substack{+22 \\ -28}$}} \% of the total gas mass is able to form stars. However, there is no clear evidence of on-going star formation, these seemingly-unstable clumps should be marginally stabilized by other forces such as magnetic fields.
\item Within the central 1.5 pc, tidal force again dominates the gravitational stability and the threshold densities quickly rise to $\ge10^{8}$ cm$^{-3}$, which strongly inhibit star formations.
\end{itemize}

\acknowledgments
We are grateful to the referee for a thorough and insightful report which helped to improve the paper.
P.-Y. H. was supported by the Ministry of Science and Technology (MoST) of Taiwan through the grants MOST 108-2112-M-001-012 and MOST 109-2112-M-001-022.  The work of W.-T. K. was supported by the grants of National Research Foundation of Korea (2019R1A2C1004857 and 2020R1A4A2002885). Jean L. Turner was supported by the NSF grant AST 2006433. This paper makes use of the following ALMA data: ADS/JAO.ALMA\#2017.1.00040.S. ALMA is a partnership of ESO (representing its member states), NSF (USA) and NINS (Japan), together with NRC (Canada), MOST and ASIAA (Taiwan), and KASI (Republic of Korea), in cooperation with the Republic of Chile. The Joint ALMA Observatory is operated by ESO, AUI/NRAO and NAOJ.
\vspace{5mm}

\facilities{ALMA}


\software{astrodendro \citep{rosolowsky08},  
          Radex \citep{radex}, 
          }

\clearpage
\appendix

Here we display the channel map of CS(7-6) line (Figure~\ref{fig-chan}), the density-distance relation for clumps with uncertainties less then 1 order of magnitude (Figure~\ref{fig-d-error}), and various physical quantities of clumps in the deprojected map (Figure~\ref{fig-4panel}).

\begin{center}
\begin{figure}[bht!]
\epsscale{1.15}
\plotone{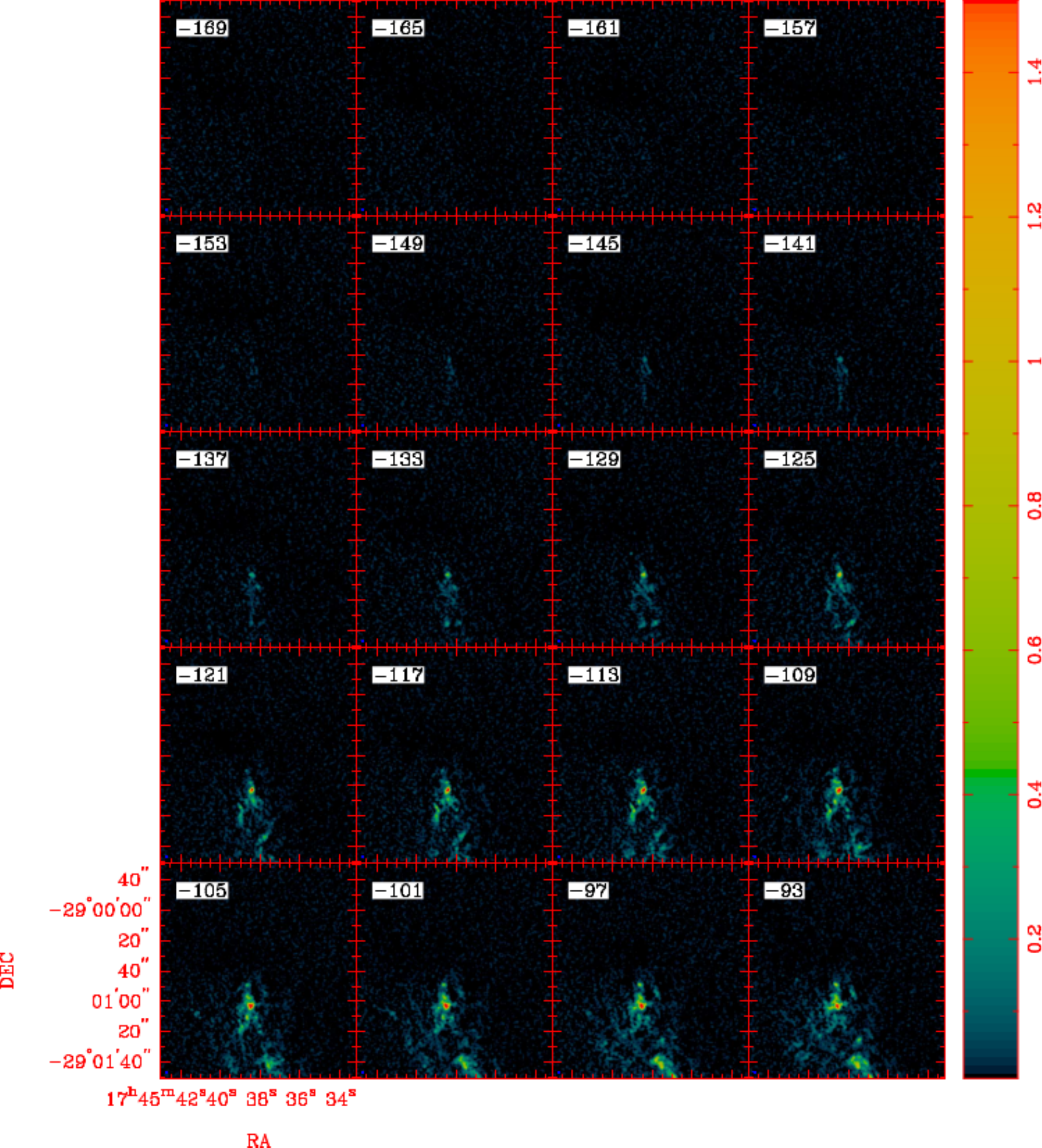}
\caption{Channel map of the CS(7-6) line. The velocity width is binned to 4 km s$^{-1}$ to reduce the numbers of panels. The LSR velocity is labeled on the top left corner in unit of km s$^{-1}$. The intensity unit of Jy beam$^{-1}$ (beam=1.3$\arcsec$).
}
\label{fig-chan}
\end{figure}
\end{center}

\addtocounter{figure}{-1}
\begin{figure}[bht!]
\begin{center}
\epsscale{1.15}
\plotone{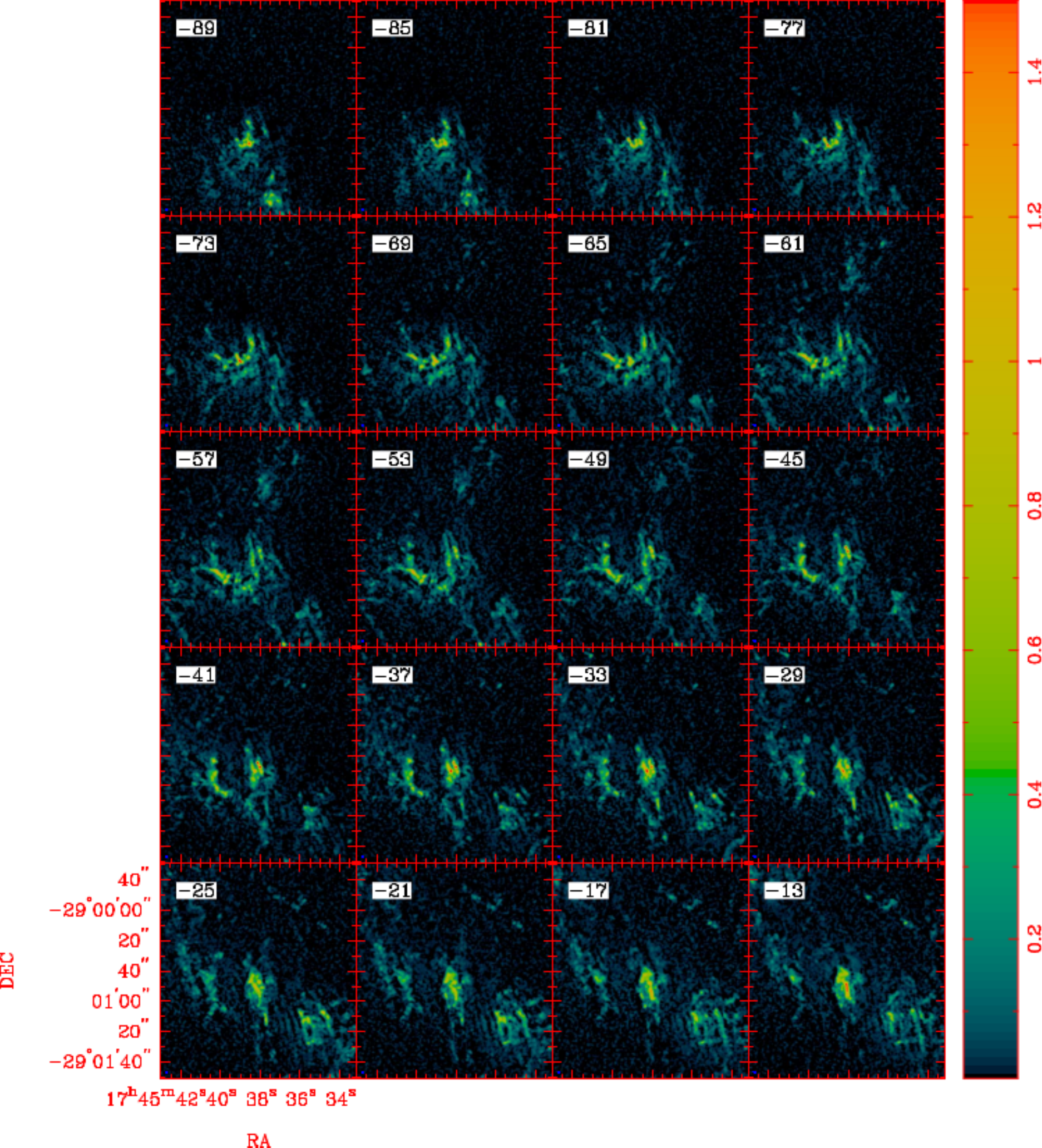}
\caption{\small Continued.}
\label{}
\end{center}
\end{figure}


\addtocounter{figure}{-1}
\begin{figure}[bht!]
\begin{center}
\epsscale{1.15}
\plotone{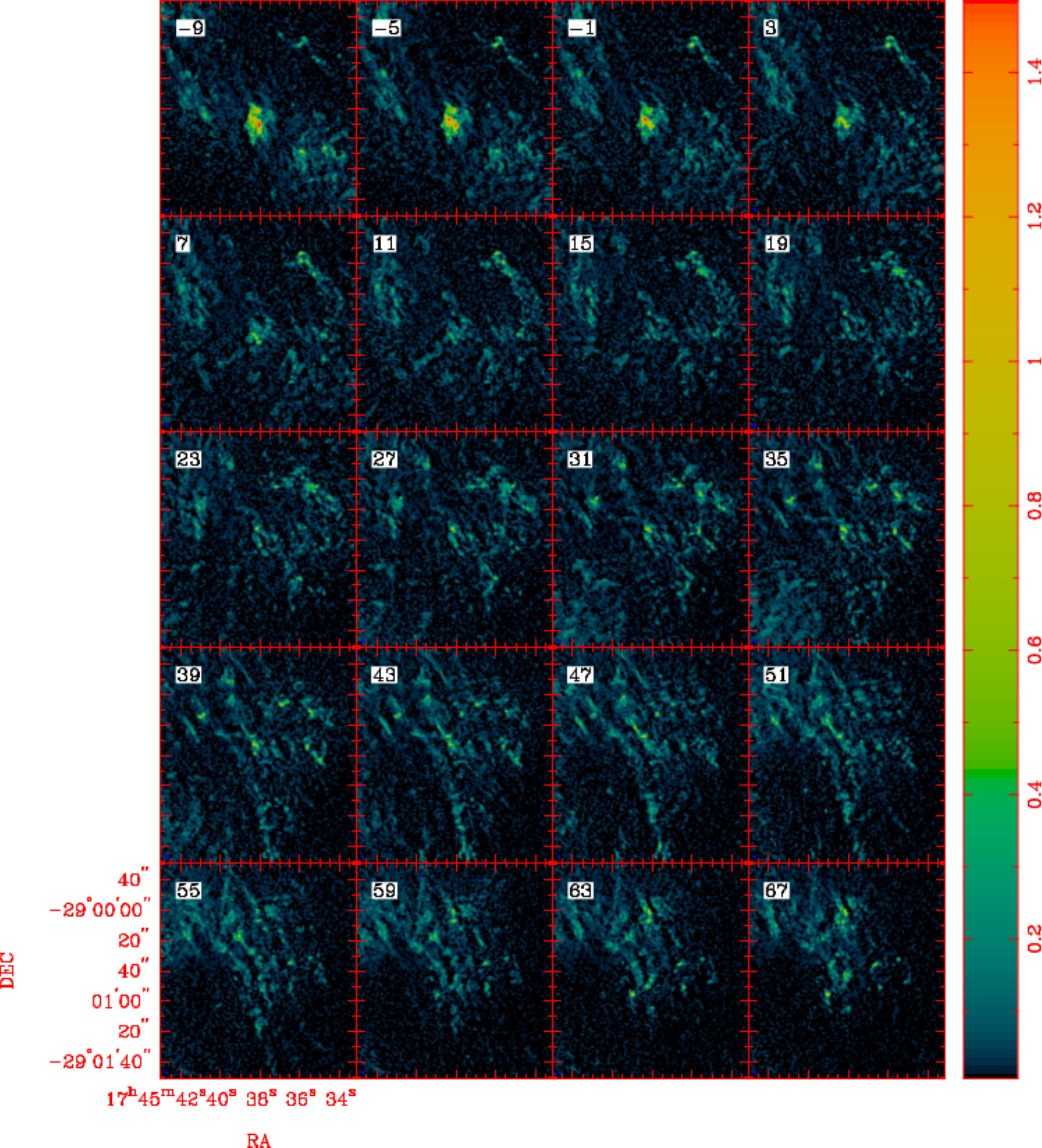}
\caption{\small Continued.}
\label{}
\end{center}
\end{figure}

\addtocounter{figure}{-1}
\begin{figure}[bht!]
\begin{center}
\epsscale{1.15}
\plotone{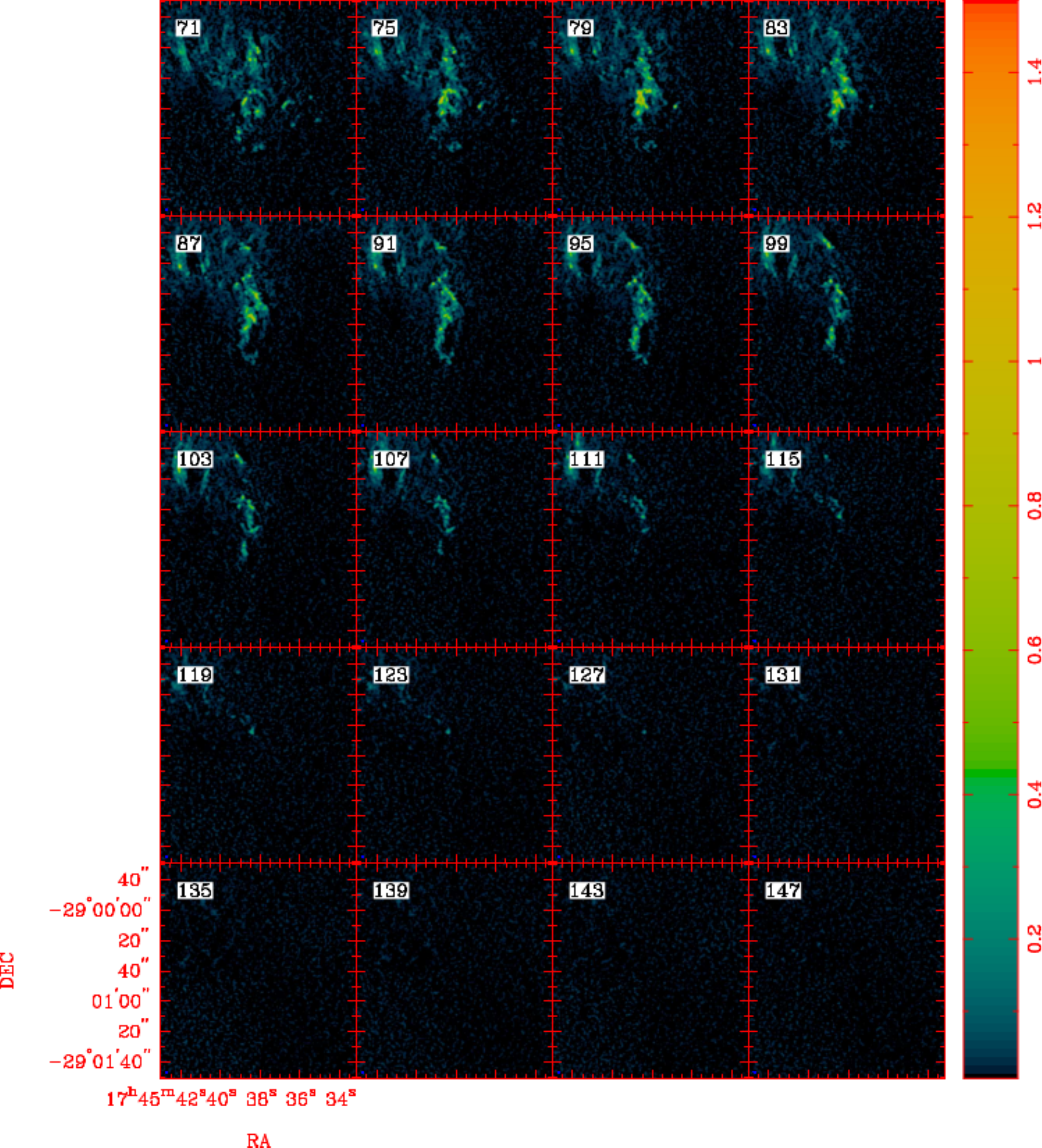}
\caption{\small Continued.}
\label{}
\end{center}
\end{figure}

\begin{figure}[bht!]
\epsscale{0.7}
\plotone{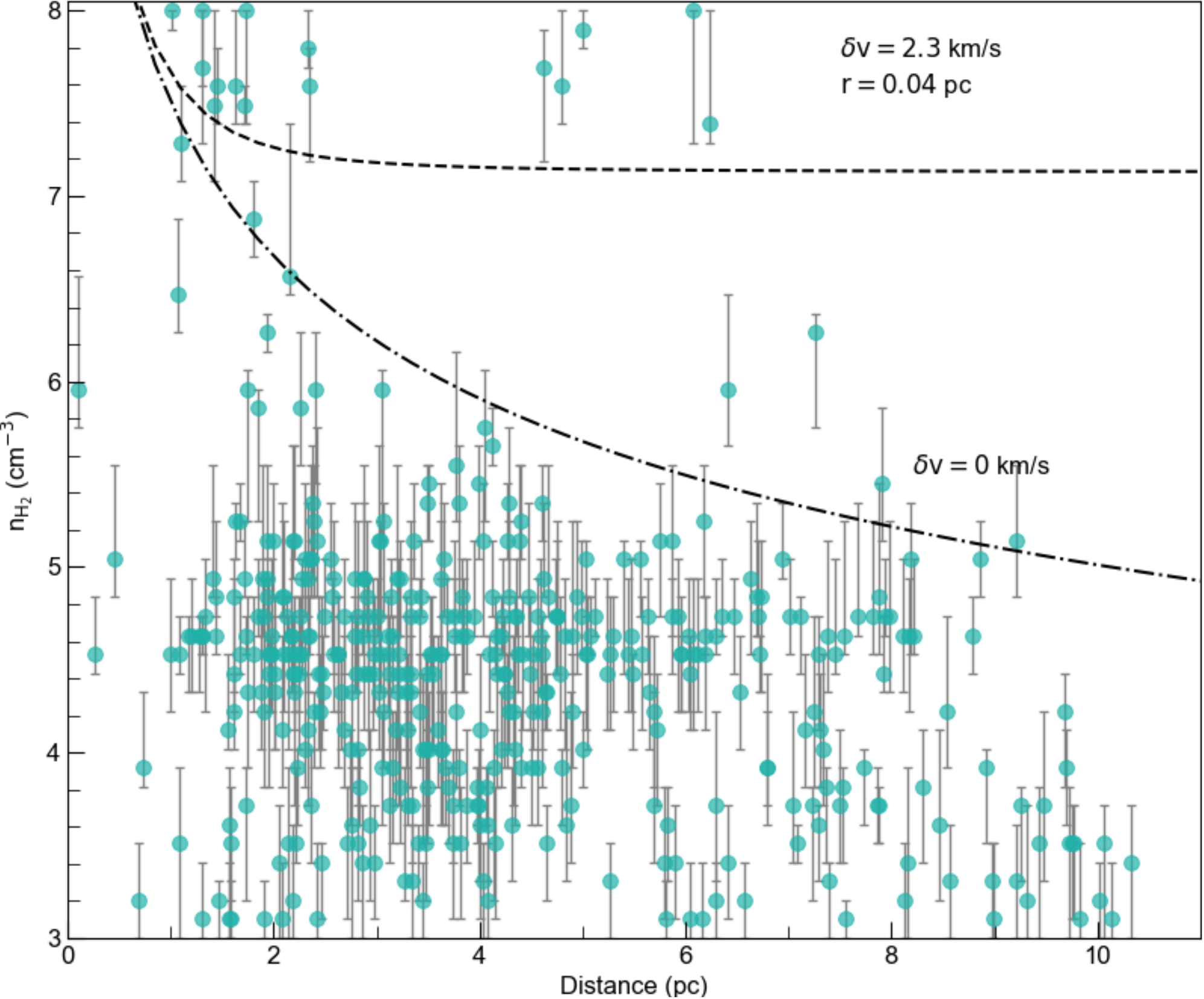}
\caption{Same as Figure~\ref{fig-deproj-den}, but the densities with uncertainties less than 1 order of magnitude are shown. Clumps with densities having upper limit ($\leq 10^{8}$ cm$^{-3}$ are also presented.)
}
\label{fig-d-error}
\end{figure}
\vspace{0.5cm}

\begin{figure*}[bht!]
\epsscale{1.15}
\plotone{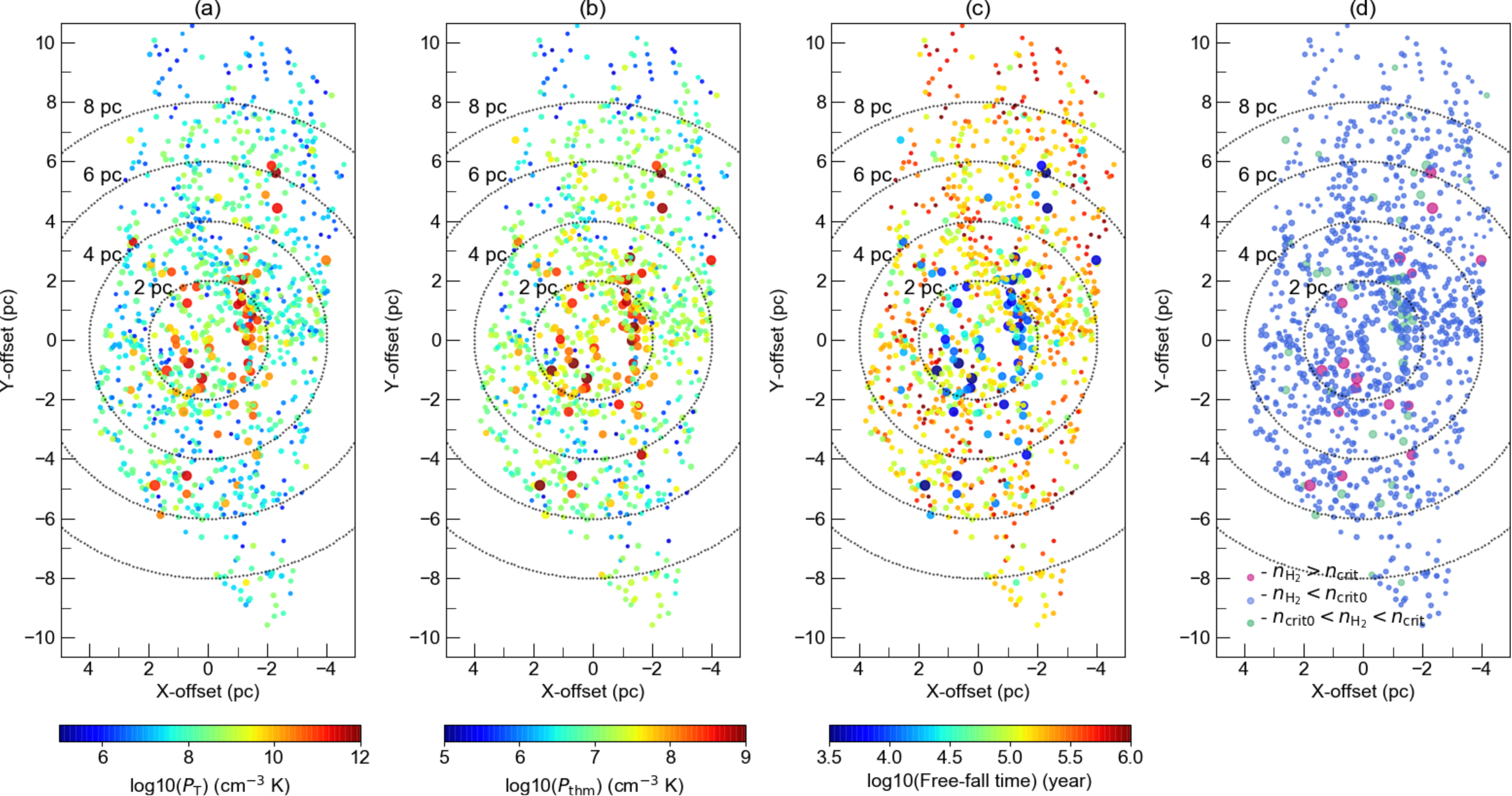}\vspace{0.4cm}
\caption{Clumps shown in the deprojected map.
(a) colors represent the turbulence pressure of clumps.
(b) colors represent the thermal pressure of clumps
(c) colors represent the free-fall time $\tau_{\rm ff}$ of clumps. The free-fall time ($\tau_{\rm ff}$) is calculated as $\tau_{\rm ff}=\sqrt[]{\frac{3\pi}{32G\rho_{gas}}}$.
(d) blue circles are clumps below $n_{\rm crit0}$ (dispersed by tidal force). Green circles are clumps within $n_{\rm crit0}$ and $n_{\rm crit}$ (stable clumps). Red circles are clumps above $n_{\rm crit}$ (able to collapse to form stars). The circle sizes are proportional to $n_{\rm H_2}$.
}
\label{fig-4panel}
\end{figure*}
\vspace{0.5cm}

\clearpage
\begin{footnotesize}
\begin{table*} [h]
\begin{center}
\caption{\em\centering\normalsize{CS Lines}}
\label{tab:line}
\begin{tabular}{cllc}
\hline\hline    \noalign {\smallskip}
Transition & Rest Frequency (GHz) & $E_\text{u}$ (K) & Critical Density (cm$^{-3}$)\\
\hline
2--1 & 97.980 & 7.1   & 3.9$\times 10^{5}$\\
3--2 & 146.969 & 14.1 & 1.2$\times 10^{6}$\\
4--3 & 195.954 & 23.5 & 2.8$\times 10^{6}$\\
5--4 & 244.935 & 35.5 & 5.5$\times 10^{6}$\\
7--6 & 344.882 & 65.8 & 1.5$\times 10^{7}$\\
\hline
\multicolumn{4}{@{}p{120mm}}{\footnotesize The collisional critical density is calculated with the Einstein coefficients and collisional rates \citep{lique} from LAMDA
data base at kinetic temperature of 60 K.}
\end{tabular}
\end{center}
\end{table*}
\end{footnotesize}

\clearpage

\begin{footnotesize}
\begin{longtable}{lcccccccc}
\caption{\em\centering{12m-Mosaic Observation}}
\label{tab:obs1}
\\\hline\hline    \noalign {\smallskip}
Line & Frequency & $\Delta f$/$\Delta v$ & Date & EB & Antennas &  Pointing & Baseline & Integration\\
    &[GHz]& [kHz/km s$^{-1}$] &  &  &   &  & [m]/[$k\lambda$] & [min]\\
(1) & (2) & (3) & (4) & (5) &  (6) & (7) & (8) & (9)\\
\hline
\multirow{2}{*}{CS(2-1)}&\multirow{2}{*}{97.980} &\multirow{2}{*}{282.3/1.0} &2017-12-24 & X317f & 43 & 68 & 15-2517/4.9--822 & 14\\
                                 &                        &&2018-06-05 & X16ec & 50 & 68 & 15-360/4.9--118 & 7\\
\hline
\multirow{1}{*}{CS(3-2)}&\multirow{1}{*}{146.969} &\multirow{1}{*}{564.5/1.3} &2018-03-15  & X1daef & 43 & 138 & 15-784/7.3--384 & 28\\
\hline

\multirow{4}{*}{CS(4-3)}&\multirow{4}{*}{195.954} &\multirow{4}{*}{564.5/0.9}&2018-08-19 &X16d2 & 43 & 150 & 15--457/9.8--298 & 34\\
                                 &                         &&2019-03-28 &X2858 & 44 & 101 &15--500/9.8--326 & 20\\
                                 &                         &&2018-12-23 &X12445 & 45 & 101 &15--500/9.8--326 & 20\\
                                 &                         &&2018-08-17 &X81da & 43 & 101 &15--500/9.8--326 & 20\\
\hline

\multirow{2}{*}{CS(5-4)}&\multirow{2}{*}{244.936} &\multirow{2}{*}{1129/1.4} &2018-03-27 & X4b71 & 43& 150& 15--784/12--640& 30 \\
                                 &                         &&2018-03-27 & X57af & 44& 150& 15--784/12--640& 30 \\
\hline
\multirow{5}{*}{CS(7-6)}&\multirow{5}{*}{342.883} &\multirow{5}{*}{1129/1.0} &2018-05-01 & X2f73& 43& 150& 15--500/17--571& 45\\
                                 &                         &&2018-04-19 & X86f1& 44& 150& 15--500/17--571& 45\\
                                 &                         &&2018-04-07 & X922b& 43& 150& 15--484/17--553& 44 \\
                                 &                         &&2018-04-07 & X78b3& 43& 150& 15--484/17--553& 45\\
                                 &                         &&2018-04-18 & X9e23& 45& 150& 15--500/17--571& 45\\
\hline\\
\multicolumn{9}{@{}p{165mm}}{\footnotesize (1)target line and the ALMA band. (2) line rest frequency. (3) default spectral resolution in frequency and velocity. (4) observing date.(5) name of ALMA execution blocks (EB). (6) number of antennas used in the observation. (7) number of pointings for mosaic observation. (8) baselines of configuration used for the observation. The min/max baseline lengths are in meter and $k\lambda$. (9) on-source integration time.}
\end{longtable}
\end{footnotesize}

\clearpage


\begin{footnotesize}
\begin{longtable}{lccccccc}
\caption{\em{7m-Mosaic Observation}}\\
\label{tab:obs2}\\
\hline\hline    \noalign {\smallskip}
Line & Frequency  & Date & EB & Antennas &  Pointing & Baseline & Integration\\
    &[GHz]&  &  &  &    & [m]/[$k\lambda$] & [min]\\
(1) & (2) & (3) & (4) & (5) &  (6) & (7) & (8)\\
\hline
\multirow{1}{*}{CS(2-1)}&\multirow{1}{*}{97.980} &2017-11-06 & X21e & 9 & 27 & 9--49/2.9--16 & 41\\
                                 &                        &2017-10-10 & X3fe1 & 9 & 27 & 9--49/2.9--16 & 41\\
\hline
\multirow{1}{*}{CS(3-2)}&\multirow{1}{*}{146.969} &2017-10-19 & X467 & 10 & 45 & 9--49/4.4--24 & 45\\
                                 &                        &2017-10-17 & X3a8 & 10 & 45 & 9--49/4.4--24 & 45\\
                                 &                        &2017-10-14 & X287 & 11 & 45 & 9--49/4.4--24 & 45\\
\hline
\multirow{1}{*}{CS(4-3)}&\multirow{1}{*}{195.954} &2018-05-02 &Xbf98 & 10 & 73 &9--48/5.9--31 & 49\\
                                 &                         &2019-05-02 &Xb438 & 10 & 73 &9--48/5.9--31 & 49\\
                                 &                         &2018-04-29 &X2a34 & 10 & 73 &9--48/5.9--31 & 49\\
                                 &                         &2018-04-29 &X1fd9 & 10 & 52 &9--49/5.9--32 & 44\\
                                 &                         &2018-04-29 &X198f & 10 & 52 &9--49/5.9--32 & 44\\
\hline
\multirow{1}{*}{CS(5-4)}&\multirow{1}{*}{244.936} &2017-10-04 & X2aa & 11& 74 & 9--49/7.3--37& 50\\
                                 &                         &2017-10-03 & X8544& 9 & 74 & 9--45/7.3--40& 50\\
                                 &                         &2017-10-03 & X80dd& 9 & 74 & 9--45/7.3--40& 50\\
                                 &                         &2017-10-23 & X179 & 10 & 79 & 9--49/7.3--37& 40\\
                                 &                         &2017-10-20 & X6aa & 10 & 79 & 9--49/7.3--37& 40\\
                                 &                         &2017-10-19 & X4c42& 10 & 79 & 9--49/7.3--37& 40\\
                                 &                         &2017-10-19 & X4981& 10 & 79 & 9--49/7.3--37& 40\\
\hline
\multirow{1}{*}{CS(7-6)}&\multirow{1}{*}{342.883} &2017-12-12 & X4d74&  10& 70 & 9--49/10.3--56& 47\\
                                 &                         &2017-10-08 & X47d7&  9& 70 & 9--49/10.3--56& 47\\
                                 &                         &2017-10-06 & X39e& 11& 70 & 9--49/10.3--56& 47\\
                                 &                         &2017-10-03 & X704& 11& 70 & 9--49/10.3--56& 47\\
                                 &                         &2018-03-11 & X355d& 10& 67 & 9--49/10.3--56& 45\\
                                 &                         &2018-03-10 & X522a& 10& 67 & 9--49/10.3--56& 45\\
                                 &                         &2018-01-21 & X69f0& 11& 67 & 9--49/10.3--56& 45\\
                                 &                         &2018-01-20 & X597e& 11& 67 & 9--49/10.3--56& 45\\
                                 &                         &2018-04-19 & X57e6& 10& 75 & 9--49/10.3--56& 38\\
                                 &                         &2018-04-11 & X12b91& 10& 75 & 9--49/10.3--56& 38\\
                                 &                         &2018-04-07 & X50f2& 11& 75 & 9--49/10.3--56& 38\\
                                 &                         &2018-04-02 & Xb7a8& 11& 75 & 9--45/10.3--56& 38\\
                                 &                         &2018-04-02 & Xb219& 11& 75 & 9--45/10.3--51& 38\\
                                 &                         &2018-04-19 & X6b30& 10& 75 & 9--49/10.3--56& 38\\
                                 &                         &2018-04-12 & X505c& 10& 75 & 9--49/10.3--56& 38\\
                                 &                         &2018-03-24 & X5a09& 11& 75 & 9--49/10.3--56& 38\\
                                 &                         &2018-03-10 & X603e& 11& 75 & 9--45/10.3--51& 38\\
                                 &                         &2018-01-21 & X5ec2& 11& 75 & 9--45/10.3--51& 38\\

\hline\\
\multicolumn{8}{@{}p{160mm}}{\footnotesize (1) target line and the ALMA band. (2) rest frequency of the line. (3) observing date.(4) name of ALMA execution blocks. (5) number of antennas used in the observation. (6) number of pointing for mosaic observation. (7) baselines of configuration used for the observation. The min/max baseline lengths are in meter and kilo-$\lambda$.  (8) on-source integration time.}

\end{longtable}
\end{footnotesize}

\clearpage
\begin{longtable}{lccccc}
\caption{\em{Total Power Observation}}\\
\label{tab:obs3}\\
\hline\hline    \noalign {\smallskip}
Line  & Frequency (GHz) & Date & EB & Antennas & Integrtion (min) \\
(1) & (2) & (3) & (4) & (5) &  (6)\\
\hline
\multirow{1}{*}{CS(2-1)}&\multirow{1}{*}{97.980} &2018-01-17 & Xd65f & 3 & 54\\
                                 &                        &2018-01-04 & X82fd & 3 & 54\\
                                 &                        &2018-01-04 & X7a47 & 3 & 54\\
\hline

\multirow{1}{*}{CS(3-2)}&\multirow{1}{*}{146.969} &2018-01-09 & X7322 & 3 & 59\\
                                 &                        &2018-01-09 & X696e & 3  & 59\\
                                 &                        &2018-01-07 & X7b59 & 3  & 59\\
                                 &                        &2018-01-04 & X8a3d & 3  & 59\\
                                 &                        &2017-12-27 & Xfc40 & 2  & 59\\
\hline

\multirow{1}{*}{CS(4-3)}&\multirow{1}{*}{195.954} &2018-06-18 &X2191 & 3 & 53\\
                                 &                         &2018-06-18 &X1c3a & 3 & 53\\
                                 &                         &2018-05-25 &Xd7b6 & 3 & 53\\
                                 &                         &2018-05-02 &Xc804 & 4 & 53\\
                                 &                         &2018-05-02 &Xbd2f & 4 & 53\\
                                 &                         &2018-05-02 &Xb37f & 4 & 53\\
                                 &                         &2018-12-11 &X71bb & 3 & 53\\
                                 &                         &2018-05-25 &Xce26 & 3 & 53\\
                                 &                         &2018-04-29 &X39bc & 2 & 53\\
                                 &                         &2018-04-29 &X2b63 & 2 & 53\\
                                 &                         &2018-04-07 &X97d5 & 3 & 38\\
                                 &                         &2018-03-27 &X6320 & 3 & 42\\

\hline

\multirow{1}{*}{CS(5-4)}&\multirow{1}{*}{244.936} &2018-03-16 &X6d6e & 3 & 52 \\
                                 &                         &2018-03-16 &X63e1 & 3 & 52 \\
                                 &                         &2018-03-13 &X53fc & 3 & 52 \\
                                 &                         &2018-03-09 &X2b3c & 3 & 52 \\
                                 &                         &2018-03-09 &X229d & 3 & 52 \\
                                 &                         &2018-01-23 &X6d48 & 3 & 52 \\
                                 &                         &2018-04-04 &X3599 & 4 & 51 \\
                                 &                         &2018-04-01 &X1f93 & 3 & 51 \\
                                 &                         &2018-03-31 &X45dd & 3 & 18 \\
                                 &                         &2018-03-31 &X35a5 & 3 & 52 \\
                                 &                         &2018-03-30 &X2821 & 3 & 48 \\
                                 &                         &2018-03-23 &X2ff4 & 3 & 51 \\
\hline

\multirow{1}{*}{CS(7-6)}&\multirow{1}{*}{342.883} &2018-04-05 &Xb461 & 4 & 53 \\
                                 &                         &2018-04-05 &Xb0ae & 4 & 13 \\
                                 &                         &2018-04-04 &X4ec9 & 2 & 53 \\
                                 &                         &2018-04-04 &X3ff0 & 4 & 53 \\
                                 &                         &2018-04-01 &X3355 & 3 & 53 \\
                                 &                         &2018-03-26 &X65bb & 3 & 53 \\
                                 &                         &2018-03-24 &X7525 & 3 & 53 \\
                                 &                         &2018-03-24 &X5b20 & 3 & 53 \\
                                 &                         &2018-05-15 &X3936 & 3 & 56 \\
                                 &                         &2018-05-13 &X25e1 & 3 & 56 \\
                                 &                         &2018-05-13 &X33d & 3 & 56 \\
                                 &                         &2018-05-09 &X58c4 & 3 & 53 \\
                                 &                         &2018-05-07 &Xedef & 3 & 53 \\
                                 &                         &2018-05-07 &Xc4d7 & 3 & 53 \\
                                 &                         &2018-05-03 &X3d00 & 4 & 47 \\
                                 &                         &2018-05-03 &X2791 & 4 & 56 \\
                                 &                         &2018-04-18 &Xa915 & 3 & 56 \\
                                 &                         &2018-05-03 &X1cb5 & 4 & 50 \\
                                 &                         &2018-04-27 &X91cd & 4 & 50 \\
                                 &                         &2018-04-20 &X10bda & 4 & 50 \\
                                 &                         &2018-04-19 &X8e40 & 4 & 50 \\
                                 &                         &2018-04-19 &X595e & 4 & 50 \\
                                 &                         &2018-04-06 &X8011 & 4 & 47 \\
                                 &                         &2018-04-06 &X5e0e & 4 & 50 \\
                                 &                         &2018-04-03 &X3bf7 & 3 & 50 \\
                                 &                         &2018-04-03 &X29a7 & 3 & 50 \\
                                 &                         &2018-08-20 &Xa5b3 & 3 & 55 \\
                                 &                         &2018-08-19 &X1058 & 3 & 55 \\
                                 &                         &2018-08-17 &X73be & 3 & 55 \\
                                 &                         &2018-06-21 &X256e & 4 & 55 \\
                                 &                         &2018-05-17 &X35ad & 3 & 55 \\
                                 &                         &2018-05-16 &Xd61f & 2 & 55 \\
                                 &                         &2018-05-14 &X2864 & 3 & 55 \\
                                 &                         &2018-05-14 &X1b7c & 3 & 55 \\
\hline\\
\multicolumn{6}{@{}p{150mm}}{\footnotesize (1)target line and the ALMA band. (2) rest frequency of the line. (3) observing date.(4) name of ALMA execution blocks. (5) number of antennas used in the observation. (6) on-source integration time.}


\end{longtable}


\clearpage

\begin{footnotesize}
\begin{table*} [h]
\begin{center}
\caption{\em\centering\normalsize{Gas mass of the 1071 clumps}}
\label{tab:mass}
\begin{tabular}{lccc}
\hline\hline    \noalign {\smallskip}
Type of clumps & Gas mass (10$^{4}$ M$_{\odot}$) & $f_{\rm gas}^{\rm (4)}$ (\%) & Number of clumps \\
\hline
(1) $n_{\rm H_2}<n_{\rm crit0}$              & $0.4${\raisebox{0.3ex}{\tiny$\substack{+0.7 \\ -0.1}$}} & $16${\raisebox{0.3ex}{\tiny$\substack{+28 \\ -7.5}$}} & 1009\\

(2) $n_{\rm crit0}<n_{\rm H_2}<n_{\rm crit}$ & $0.7${\raisebox{0.3ex}{\tiny$\substack{+0.6 \\ -0.1}$}} & $28${\raisebox{0.3ex}{\tiny$\substack{+24 \\ -12}$}} & 46\\

(3) $n_{\rm H_2}>n_{\rm crit}$               & $1.4${\raisebox{0.3ex}{\tiny$\substack{+0.5 \\ -0.4}$}} &
$56${\raisebox{0.3ex}{\tiny$\substack{+22 \\ -28}$}} & 16\\
\hline
\multicolumn{4}{@{}p{100mm}}{\footnotesize (1) clumps with $n_{\rm H_2}$ lower than tidal threshold, (2) clumps gravitationally stable, (3) clumps are able to collapse to form stars, (4) $f_{\rm gas}$ is the fraction of total gas mass. The total gas mass is $2.5${\raisebox{0.3ex}{\tiny$\substack{+1.0 \\ -0.4}$}} $\times10^{4}$ M$_{\odot}$.}
\end{tabular}
\end{center}
\end{table*}
\end{footnotesize}


\begin{footnotesize}
\begin{table*} [h]
\begin{center}
\caption{\em\centering\normalsize{Clumps above $n_\text{crit}$}}
\label{tab:sfcloud}
\begin{tabular}{lcccccccc}
\hline\hline    \noalign {\smallskip}
ID & RA. (J2000) & Dec. (J2000) & $R_\text{s}$ & $V_\text{cen}$ & $\sigma_\text{vden}$& $T_\text{k}$ & $n_{H2}$ & $M_\text{gas}$\\
    &   &  & (pc) & (km s$^{-1}$) & (km s$^{-1}$)  &  (K)  & (cm$^{-3}$) & (M$_{\odot}$)\\
(1) & (2) & (3) & (4) & (5) &  (6) & (7) & (8) & (9)\\
\hline
446& 17:45:40.40 & -29:01:13.04 & 0.036$\pm$0.009 & -105.1 & 1.5$\pm$0.6 & 20$_{-10}^{+500}$ & 7.18$_{-2.45}^{+0.82}$ & 206$_{-205}^{+1140}$\\
933& 17:45:37.40 & -29:00:50.34 & 0.022$\pm$0.006 & -60.8 & 1.6$\pm$0.8 & 10$_{-7.27}^{+10}$ & 8.00$_{-1.22}^{+0.00}$ & 291$_{-273}^{+0}$\\
974& 17:45:41.40 & -29:01:20.52 & 0.023$\pm$0.009 & -59.4 & 0.9$\pm$0.4 & 20$_{-10}^{+230}$ & 7.49$_{-2.14}^{+0.51}$ & 109$_{-108}^{+244}$\\
1076& 17:45:40.83 & -29:00:56.52 & 0.026$\pm$0.007 & -49.0 & 1.3$\pm$0.5 & 10$_{-7.27}^{+10}$ & 7.59$_{-0.41}^{+0.20}$ & 196$_{-120}^{+118}$\\
1272& 17:45:35.58 & -29:01:58.18 & 0.025$\pm$0.007 & -34.7 & 1.2$\pm$0.6 & 10$_{-7.27}^{+10}$ & 7.59$_{-0.20}^{+0.41}$ & 165$_{-62}^{+258}$\\
1281& 17:45:34.85 & -29:01:07.11 & 0.051$\pm$0.004 & -29.8 & 3.8$\pm$0.5 & 10$_{-7.27}^{+10}$ & 8.00$_{-0.71}^{+0.00}$ & 3693$_{-2980}^{+0}$\\
1422& 17:45:35.58 & -29:01:12.27 & 0.038$\pm$0.005 & -26.0 & 2.3$\pm$0.4 & 10$_{-7.27}^{+10}$ & 7.90$_{-0.10}^{+0.10}$ & 1268$_{-265}^{+336}$\\
1709& 17:45:44.41 & -28:59:59.25 & 0.036$\pm$0.007 & -16.0 & 1.5$\pm$0.4 & 10$_{-7.27}^{+110}$ & 8.00$_{-2.76}^{+0.00}$ & 1267$_{-1265}^{+0}$\\
2251& 17:45:41.01 & -29:00:27.12 & 0.030$\pm$0.004 & 7.2 & 2.7$\pm$0.5 & 10$_{-7.27}^{+10}$ & 8.00$_{-0.71}^{+0.00}$ & 742$_{-599}^{+0}$\\
2565& 17:45:41.65 & -28:59:57.19 & 0.026$\pm$0.007 & 17.3 & 1.6$\pm$0.5 & 10$_{-7.27}^{+10}$ & 8.00$_{-0.61}^{+0.00}$ & 502$_{-380}^{+0}$\\
2851& 17:45:41.24 & -29:00:27.12 & 0.034$\pm$0.006 & 27.3 & 1.9$\pm$0.5 & 10$_{-7.27}^{+10}$ & 7.59$_{-0.20}^{+0.41}$ & 437$_{-164}^{+681}$\\
3044& 17:45:40.99 & -29:00:14.48 & 0.030$\pm$0.005 & 35.5 & 2.6$\pm$0.6 & 10$_{-7.27}^{+10}$ & 8.00$_{-0.10}^{+0.00}$ & 798$_{-167}^{+0}$\\
3440& 17:45:43.46 & -29:00:24.79 & 0.049$\pm$0.007 & 50.2 & 3.2$\pm$0.4 & 10$_{-7.27}^{+10}$ & 7.69$_{-0.51}^{+0.20}$ & 1601$_{-1106}^{+960}$\\
3446& 17:45:39.71 & -29:00:07.25 & 0.058$\pm$0.005 & 51.5 & 3.3$\pm$0.4 & 10$_{-7.27}^{+10}$ & 7.49$_{-0.41}^{+0.51}$ & 1704$_{-1038}^{+3809}$\\
3581& 17:45:37.46 & -29:01:01.94 & 0.044$\pm$0.010 & 51.7 & 1.6$\pm$0.3 & 20$_{-10}^{+500}$ & 7.29$_{-2.14}^{+0.61}$ & 449$_{-446}^{+1390}$\\
3659& 17:45:42.15 & -29:00:15.51 & 0.044$\pm$0.009 & 53.3 & 1.0$\pm$0.1 & 10$_{-7.27}^{+500}$ & 7.49$_{-3.06}^{+0.51}$ & 719$_{-718}^{+1607}$\\
\hline
\multicolumn{9}{@{}p{160mm}}{\footnotesize (1) the ID of clumps, (2), (3) the positions of clumps, (4) the effective radius ($R_\text{s}$) of clumps, (5) the central velocities of clumps, (6) the velocity dispersion derived in astrodendro, (7) kinetic temperature, (8) volume density of H$_{2}$, (9) gas mass. Note that 5 out of them have large uncertainties of $T_\text{k}$ and $n_H2$ (ID=446, 974, 1079, 3581, 3659).}
\end{tabular}
\end{center}
\end{table*}
\end{footnotesize}


\bibliographystyle{aasjournal}


\end{document}